\documentclass[twocolumn]{aastex631}

\usepackage{amsmath,amssymb,amstext}
\usepackage{gensymb}
\usepackage{multirow}
\usepackage{rotating}
\usepackage{tabularx}
\usepackage{comment}
\usepackage[version=4]{mhchem} 

\defcitealias{LustigYaeger2023}{Lustig-Yaeger \& Fu et al. 2023}
\defcitealias{Moran2023}{Moran \& Stevenson et al. 2023}
\defcitealias{May23}{May \& MacDonald et al. 2023}
\defcitealias{Kirk2024}{Kirk et al. 2024}

\begin{document}

\title{JWST COMPASS: The first near- to mid-infrared transmission spectrum of the hot super-Earth L\,168-9\,b}

\author[0000-0003-4157-832X]{Munazza K. Alam}
\affiliation{Space Telescope Science Institute, 3700 San Martin Drive, Baltimore, MD 21218, USA}
\affiliation{Carnegie Science Earth and Planets Laboratory, 5241 Broad Branch Road, NW, Washington, DC 20015, USA}

\author[0000-0002-8518-9601]{Peter Gao} 
\affiliation{Carnegie Science Earth and Planets Laboratory, 5241 Broad Branch Road, NW, Washington, DC 20015, USA}

\author[0000-0002-4489-3168]{Jea Adams Redai} 
\affiliation{Center for Astrophysics ${\rm \mid}$ Harvard {\rm \&} Smithsonian, 60 Garden St, Cambridge, MA 02138, USA}

\author[0000-0003-0354-0187]{Nicole L. Wallack}
\affiliation{Carnegie Science Earth and Planets Laboratory, 5241 Broad Branch Road, NW, Washington, DC 20015, USA}

\author[0000-0002-0413-3308]{Nicholas F. Wogan}
\affiliation{Space Science Division, NASA Ames Research Center, Moffett Field, CA 94035}

\author[0000-0002-8949-5956]{Artyom Aguichine}
\affiliation{Department of Astronomy and Astrophysics, University of California, Santa Cruz, CA 95064, USA}

\author[0000-0002-1092-2995]{Anne Dattilo}
\affiliation{Department of Astronomy and Astrophysics, University of California, Santa Cruz, CA 95064, USA}

\author[0000-0001-8703-7751]{Lili Alderson} 
\affiliation{School of Physics, University of Bristol, HH Wills Physics Laboratory, Tyndall Avenue, Bristol BS8 1TL, UK}

\author[0000-0003-1240-6844]{Natasha E. Batalha}
\affiliation{NASA Ames Research Center, Moffett Field, CA 94035, USA}

\author[0000-0002-7030-9519]{Natalie M. Batalha}
\affiliation{Department of Astronomy and Astrophysics, University of California, Santa Cruz, CA 95064, USA}



\author[0000-0002-4207-6615]{James Kirk}
\affiliation{Department of Physics, Imperial College London, Prince Consort Road, London, SW7 2AZ, UK}

\author[0000-0003-3204-8183]{Mercedes L\'opez-Morales} 
\affiliation{Center for Astrophysics ${\rm \mid}$ Harvard {\rm \&} Smithsonian, 60 Garden St, Cambridge, MA 02138, USA}


\author[0000-0002-7500-7173]{Annabella Meech}
\affiliation{Center for Astrophysics ${\rm \mid}$ Harvard {\rm \&} Smithsonian, 60 Garden St, Cambridge, MA 02138, USA}

\author[0000-0002-6721-3284]{Sarah E. Moran} 
\affiliation{Department of Planetary Sciences and Lunar and Planetary Laboratory, University of Arizona, Tuscon, AZ, USA}


\author[0009-0008-2801-5040]{Johanna Teske} 
\affiliation{Earth and Planets Laboratory, Carnegie Institution for Science, 5241 Broad Branch Road, NW, Washington, DC 20015, USA}
\affiliation{The Observatories of the Carnegie Institution for Science, 813 Santa Barbara St., Pasadena, CA 91101, USA}

\author[0000-0003-4328-3867]{Hannah R. Wakeford} 
\affiliation{School of Physics, University of Bristol, HH Wills Physics Laboratory, Tyndall Avenue, Bristol BS8 1TL, UK}

\author[0000-0003-2862-6278]{Angie Wolfgang}
\affiliation{Eureka Scientific Inc., 2452 Delmer Street Suite 100, Oakland, CA 94602-3017}

\begin{abstract}

We present the first broadband near- to mid-infrared (3--12\,$\mu$m) transmission spectrum of the highly-irradiated ($\rm T_{eq}$= 981\,K) M dwarf rocky planet L\,168-9\,b (TOI-134\,b) observed with the NIRSpec and MIRI instruments aboard JWST. We measure the near-infrared transit depths to a combined median precision of 20 ppm across the three visits in 54 spectroscopic channels with uniform widths of 60 pixels ($\sim$0.2\,$\mu$m wide; R$\sim$100), and the mid-infrared transit depths to 61 ppm median precision in 48 wavelength bins ($\sim$0.15\,$\mu$m wide; R$\sim$50). We compare the transmission spectrum of L\,168-9\,b to a grid of 1D thermochemical equilibrium forward models, and rule out atmospheric metallicities of less than 100$\times$ solar (mean molecular weights $<$4 g mol$^{-1}$) to 3$\sigma$ confidence assuming high surface pressure ($>$1 bar), cloudless atmospheres. Based on photoevaporation models for L\,168-9\,b with initial atmospheric mass fractions ranging from 2--100\%, we find that this planet could not have retained a primordial H/He atmosphere beyond the first 200 Myr of its lifetime. Follow-up MIRI eclipse observations at 15\,$\mu$m could make it possible to confidently identify a CO$_{2}$-dominated atmosphere on this planet if one exists.   

\end{abstract}


\section{Introduction} 
\label{sec:intro}

Atmospheric characterization of terrestrial, and potentially habitable, exoplanets is a key endeavor driving the focus of next-generation space-based facilities such as the Habitable Worlds Observatory (HWO; e.g., \citealt{Checlair21,Hall23,Vaughan23,Borges24}), the Large Interferometer for Exoplanets mission (LIFE; e.g., \citealt{Defrere18,Quanz21,Quanz22,Matsuo23}), and ground-based extremely large telescopes (ELTs; e.g., \citealt{Snellen15,Hawker19,Leung20,Currie23,Zhang24}). Although these future efforts are directed toward high-contrast imaging of terrestrial planets and detecting their atmospheres, progress toward these goals can already be made using transiting systems. Transiting terrestrial planets in nearby M dwarf systems in particular offer the most favorable prospects for spectroscopic observations due to their large planet-to-star radius ratios, frequent transit events, and close-in habitable zones (e.g., \citealt{Dressing15}). However, intense X-ray and extreme-UV irradiation from M dwarf host stars can result in significant atmospheric mass loss ($\sim$$10^{8}$--$10^{10}$\,$\rm g\,s^{-1}$; \citealt{DosSantos23}), thus influencing the persistence of substantial primordial  (e.g., \citealt{Kasting83,Owen12,Rogers15,Airapetian17,Rogers21}) or secondary \citep{Dong18,Kite20} atmospheres on rocky planets over their lifetimes. 

Spectroscopic observations of small (R\,$\lesssim$\,1.7\,$\rm R_{\oplus}$) planets orbiting nearby low-mass M dwarfs have not yet provided definitive evidence of the presence of a terrestrial planet atmosphere. Transmission spectroscopy of small planets with Hubble's Widefield Camera 3 (WFC3; e.g., \citealt{DeWit16,DeWit18,Wakeford19,Damiano22}) and large ground-based telescopes (e.g. \citealt{Diamond-Lowe18,Diamond-Lowe20,Diamond-Lowe23}) have ruled out thick hydrogen/helium-dominated atmospheres for these planets. Higher mean molecular weight secondary atmospheres are possible \citep{Moran18}, but the precision and resolution of these observations cannot explore this scenario.

With JWST's exquisite precision, we can now begin probing the presence of secondary atmospheres on rocky planets. However, JWST transmission observations thus far have been challenging, since the predicted amplitude of atmospheric spectral features ($\lesssim$20\,ppm) is comparable to the expected pre-launch ($\sim$10-20\,ppm; \citealt{Greene16,Rustamkulov22}) and measured ($\sim$5\,ppm; \citetalias{LustigYaeger2023}) instrument noise floors. To date, observations of rocky planets with JWST NIRSpec/G395H have yielded featureless transmission spectra ruling out atmospheres dominated by H/He, methane, or water \citepalias{LustigYaeger2023}. However, interpretations of these spectra have been muddled by the influence of stellar activity \citepalias{Moran2023}, visit-to-visit variability \citepalias{May23}, offsets between the NRS1 and NRS2 detectors, \citep{Damiano24,Wallack2024}, differences in data reductions \citepalias{Kirk2024}, and lower than expected transit depth precisions \citep{Alderson2024}. These astrophysical and instrumental systematics are clouding our holistic understanding of the atmospheric properties of small planets.    

To better understand the prevalence and properties of small exoplanet atmospheres at the population level, the JWST COMPASS (Compositions of Mini-Planet Atmospheres for Statistical Study) program (GO 2512; PIs: Batalha \& Teske) is obtaining NIRSpec/G395H transmission spectra for a sample of 11 quantitatively-selected 1--3 $\rm R_{\oplus}$ planets -- including six targets with radii smaller than 1.7~${\rm R_{\earth}}$. The program's full statistical sample will include the addition of a target from GTO 1224. COMPASS aims to build a link between planetary demographics and atmospheric characterization of super-Earths and sub-Neptunes by exploring the detectability of small planet atmospheres and their compositional diversity -- thereby enabling population-level atmospheric constraints \citep{Batalha23}. 
The focus of this study is one of the smallest and hottest COMPASS targets: L\,168-9\,b (TOI-134\,b; \citealt{Astudillo-Defru20}). L\,168-9\,b is a highly-irradiated ($\rm T_{eq}$=981\,K) planet, 
with a radius of 1.39\,$\rm R_{\oplus}$ and a mass of 
 4.6\,$\rm M_{\oplus}$,
 in a 1.4\,d orbital period around a bright ($K$=7.1) M1V star (see Table \ref{tab:sys_params}). 

Here we present the 3--12\,$\mu$m transmission spectrum of L\,168-9\,b using JWST NIRSpec/G395H observations obtained as part of the COMPASS program and MIRI/LRS data taken during commissioning (Program COM 1033; PI: Kendrew). The structure of this paper is as follows: We describe our observations in \S \ref{sec:obs} and detail our data reduction techniques and light curve fitting methods in \S \ref{sec:data_reduction}. In  \S \ref{sec:results}, we present the near- to mid-infrared transmission spectrum of L\,168-9\,b, and interpret our results using 1D thermochemical equilibrium forward models. We place L\,168-9\,b in the context of other rocky exoplanets, investigate its atmospheric loss history, and discuss future characterization opportunities in \S \ref{sec:discussion}. We summarize our results in \S \ref{sec:summary}.  

\section{Observations} 
\label{sec:obs}

\begin{deluxetable}{cc}

\tablewidth{0pt}
\tablehead{\colhead{Stellar Parameters}  & \colhead{} } %
\startdata
K (mag)  &  7.082 $\pm$ 0.03\tablenotemark{a} \\
$M_{\star}$ ($R_\odot$)& 0.62  $\pm$ 0.03\tablenotemark{a}\\
$R_{\star}$ ($R_\odot$)& 0.600 $\pm$ 0.022\tablenotemark{a}\\
$T_{\rm{eff}}$  (K)& 3800 $\pm$ 70\tablenotemark{a}\\
log(g) (cgs) & 4.04  $\pm$ 0.49\tablenotemark{a}\\
$[$Fe/H$]_{\star}$ & 0.04 $\pm$ 0.17\tablenotemark{a} \\
Age (Gyr) & 2.99 $\pm$ 0.64\tablenotemark{b} \\
\cutinhead{Planetary Parameters}
Period (days)&  1.40150 $\pm$ 0.00018\tablenotemark{a} \\
M$_\mathrm{P}$ (M$_{\earth}$)& 4.60 $\pm$ 0.56\tablenotemark{a} \\
R$_\mathrm{P}$ (R$_{\earth}$)& 1.39 $\pm$ 0.09\tablenotemark{a}\\
T$_\mathrm{eq}$ (K)&  980 $\pm$ 20\tablenotemark{a} \\
Semi-major axis (AU) & 0.02091 $\pm$ 0.00024\tablenotemark{a}   
\tablecaption{System parameters for L\,168-9\,b.}
\enddata 
\label{tab:sys_params}
\tablenotetext{a}{\cite{Astudillo-Defru20} and references therein}
\tablenotetext{b}{\cite{Engle2023}}
\end{deluxetable}

\subsection{NIRSpec}

We observed three transits of L\,168-9\,b with JWST NIRSpec (Near-InfraRed Spectrograph; \citealt{Jakobsen22,Birkmann22}) on UT 03 June 2023, 18 June 2023, and 27 June 2023 using the high-resolution (R$\sim$2700) G395H mode, which provides spectroscopy between 2.87--5.14\,$\mu$m across the NRS1 and NRS2 detectors (with a $\sim$0.1\,$\mu$m detector gap between 3.72--3.82\,$\mu$m). The observations were taken with the NIRSpec Bright Object Time Series (BOTS) mode using the SUB2048 subarray, the F290LP filter, the S1600A1 slit, and the NRSRAPID readout pattern.  Each 5.97\,hr exposure consisted of 3359 integrations (divided into three segments) with four groups per integration, and was designed to be centered on the 1.3\,hr transit event with sufficient out-of-transit baseline. The first visit suffered from a high gain antenna (HGA) move during UT 03 June 2023 around $\sim$MJD 60098.45027, affecting integrations $\sim$1386-1392. 

\subsection{MIRI}

A single transit of L\,168-9\,b was observed with JWST MIRI (Mid-InfraRed Instrument; \citealt{Rieke15a,Rieke15b}) on UT 29 May 2022 as part of commissioning activities to characterize the performance of the slitless Low Resolution Spectroscopy (LRS; \citealt{Kendrew15,Kendrew18}) mode for time-series observations \citep{Bouwman23}. The observations were taken over a 4.14\,hr exposure consisting of 9371 integrations with nine groups per integration and divided into five segments. The exposure was taken with the SLITLESSPRISM subarray, the F1000W filter, the FASTR1 readout pattern, and designed to cover sufficient pre- and post-transit baseline, detector settling time ($\sim$30 minutes), and flexibility in scheduling. An HGA move occurred during the MIRI transit during integrations $\sim$2878-2885 ($\sim$MJD 59728.39444).

\section{Data Reduction} 
\label{sec:data_reduction}

We reduced our observations of L\,168-9\,b using three independent data reduction pipelines, which are described below: {\tt Aesop} (\S \ref{sec:aesop}), {\tt Tiberius} (\S \ref{sec:tiberius}), and {\tt Eureka!} (\S \ref{sec:eureka}). 

\begin{figure*}
\begin{centering}
\includegraphics[width=0.99\textwidth]{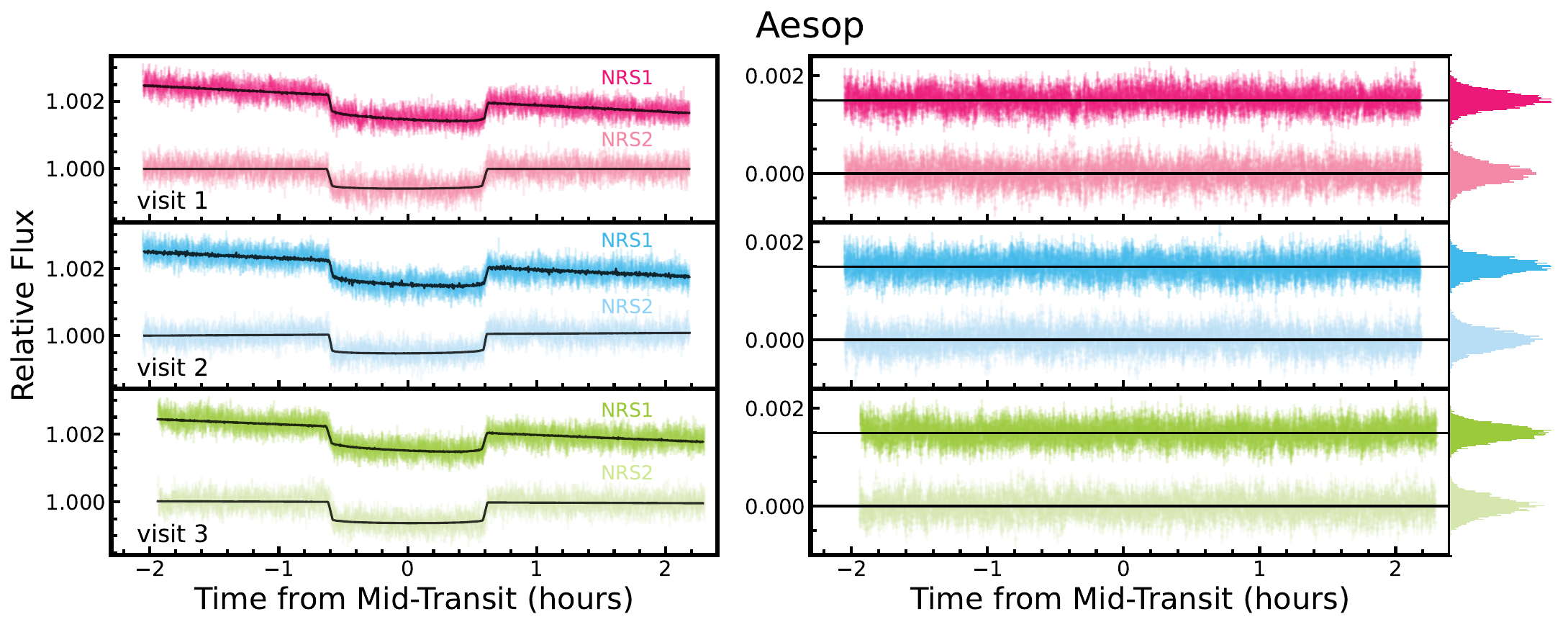}
\includegraphics[width=0.99\textwidth]{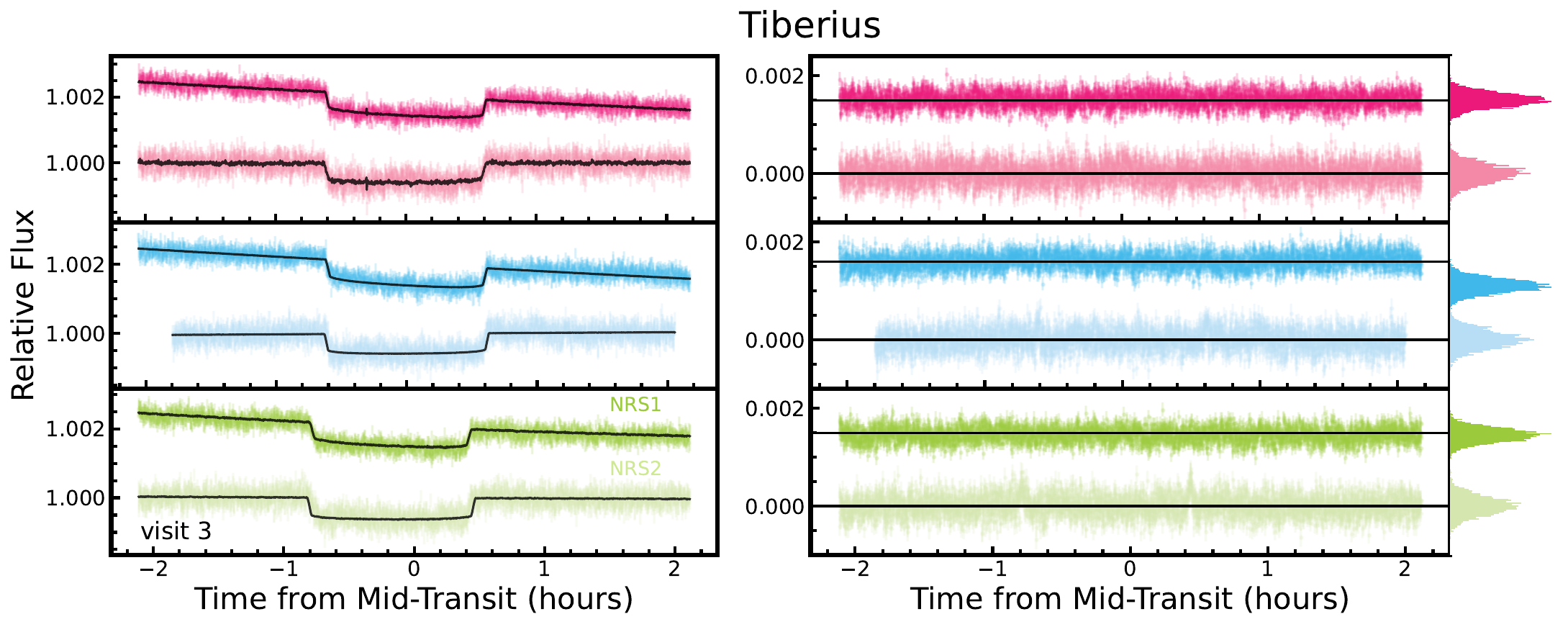}
\includegraphics[width=0.99\textwidth]{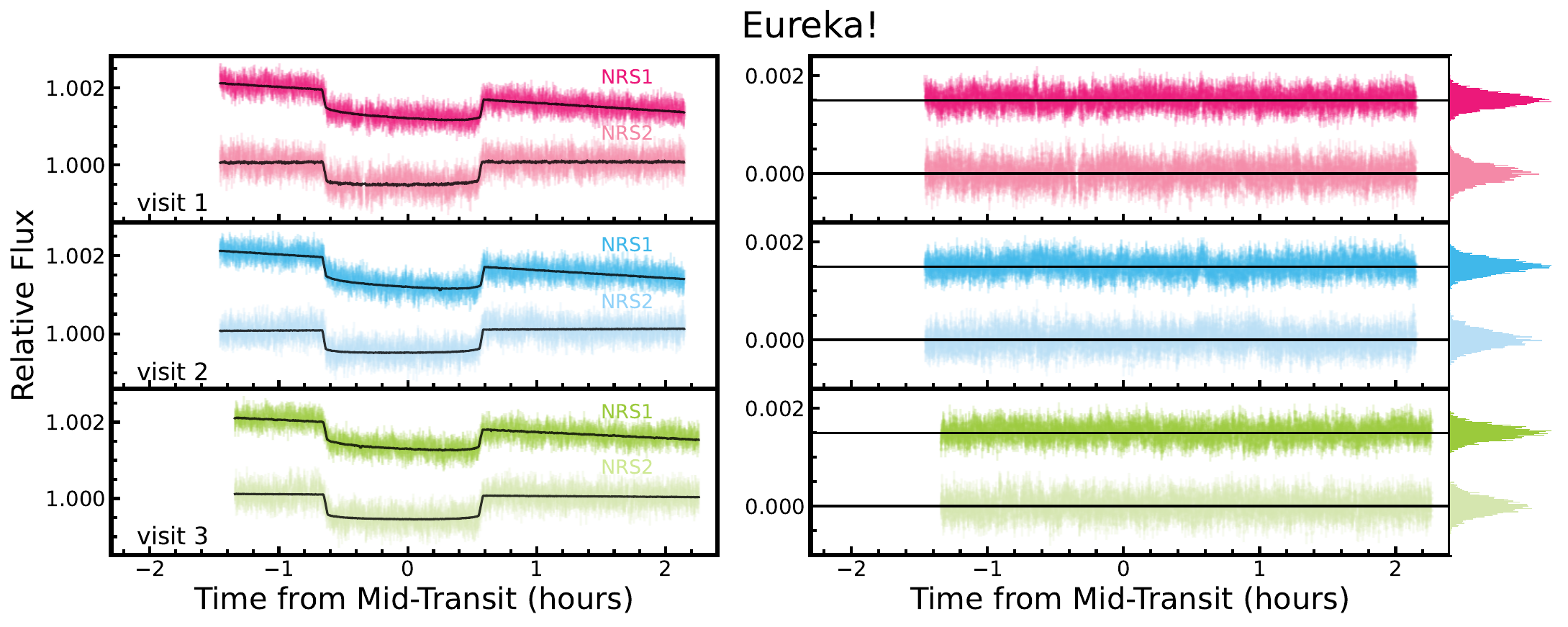}
\caption{The NIRSpec/G395H white light curves (left) for L\,168-9\,b from the {\tt Aesop} (top), {\tt Tiberius} (middle), and {\tt Eureka!} (bottom) reductions for visits 1 (pink), 2 (blue), and 3 (green), compared to the best-fitting transit+systematics models (black). The residuals for each visit are shown in the right panel. } 
\label{fig:wlc}   
\end{centering}
\end{figure*} 

\subsection{NIRSpec G395H}

\subsubsection{Aesop}
\label{sec:aesop}

{\tt Aesop} is a JWST data reduction pipeline developed for the reduction and analysis of NIRSpec G395H observations that has been benchmarked with several other pipelines in the literature (see e.g., \citealt{Alderson2023}). In our analysis, we treated the data sets for the NRS1 and NRS2 detectors, as well as for each visit, separately.  Starting with the raw uncalibrated JWST data products, we first ran the standard Stage 1 steps of the STScI {\tt jwst} pipeline (version 1.8.2, context map 1078; \citealt{Bushouse2022}) for time-series observations, including corrections for saturation, bias, linearity, and dark current. We set the detection threshold for the jump step to 15, applied a custom group-level background subtraction to remove 1/$f$ noise using a 15$\sigma$ threshold and a second-order polynomial, and performed standard ramp fitting. We then used the standard steps for Stage 2 of the {\tt jwst} pipeline to extract the integration exposure times and the 2D wavelength map. 

We performed additional cleaning steps to replace poor data quality pixels (flagged as saturated, hot, dead, do not use, low quantum efficiency, or no gain value) with the median of the neighboring pixels and removed residual 1/$f$ noise by calculating and subtracting the median pixel value of each column. To extract the 1D time-series stellar spectra, we fit a Gaussian profile to each column of a given integration to find the center of the spectral trace and smoothed the trace centers with a median filter, and then we fit a fourth-order polynomial to the smoothed trace centers. We extracted the 1D stellar spectra by summing up the flux within a 10-pixel wide aperture, calculating the uncertainties in the stellar spectra assuming photon noise. We obtained the shifts in the $x$ and $y$ pixel positions throughout the observation by cross-correlating the 1D stellar spectra to the first spectrum in the time-series. 

We generated the broadband (white) light curves by summing up the stellar flux in the time-series between 2.862704–3.714356\,$\mu$m for NRS1 and 3.819918–5.082485\,$\mu$m for NRS2. For the spectroscopic light curves, we generated 54 binned light curves with uniform widths of 60 pixels ($\sim$0.2\,$\mu$m wide) based on the G395H wavelength solution.  For both our white and spectroscopic light curves, we iteratively removed data points that were greater than 3$\sigma$ outliers in the residuals and also re-scaled our flux uncertainties following \citet{Pont06}. We fit the broadband and spectroscopic light curves for NRS1 and NRS2 using a least-squares minimizer. We fit each light curve with a two-component function consisting of a transit model generated using {\tt batman} \citep{Kreidberg2015} multiplied by a systematics model of the form:

\begin{equation}
    S(\lambda) =  c_{0} + (c_{1} \times x) + (c_{2} \times y) + (c_{3} \times t)
\end{equation}

\noindent where  $x$ and $y$ are the $x$- and $y$- pixel positional shifts on the detector, $t$ is time, and  the coefficients $c_{0}$, $c_{1}$, and $c_{2}$ are constants to be fit.

We first fit the broadband NRS1 and NRS2 white light curves for each visit by fixing the period, $P$, to the value in \citet{Astudillo-Defru20}, setting the eccentricity, $e$, to zero and argument of periastron, $\omega$, to ninety, and fitting for the mid-transit time, $T_{0}$, the scaled semi-major axis, $a/R_{\star}$, inclination, $i$,  planet-to-star radius ratio, $R_{p}/R_{\star}$, and stellar baseline flux using wide uniform priors. For the spectroscopic light curves, we fixed $T_{0}$, $a/R_{\star}$, and $i$ to the best-fit values from the broadband light curves (presented in Table \ref{tab:wlc_fits}) and fit for $R_{p}/R_{\star}$. We adopted the optimized parameters and their standard deviations as the best-fitting value and associated uncertainty for our fitted parameters. We held the non-linear limb-darkening coefficients \citep{Claret00} fixed to theoretical values, which we computed using {\tt ExoTiC-LD} \citep{ExoTiC-LD} with Set One of the MPS-ATLAS stellar models \citep{Kostogryz2022,Kostogryz2023} and the stellar effective temperature $T_{\rm{eff}}$, surface gravity log(g), and metallicity $\rm [Fe/H]_{\star}$ values given in Table \ref{tab:sys_params}. The {\tt Aesop} best-fit white light curves and residuals for each visit are shown in the top panel of Figure \ref{fig:wlc}.   

\subsubsection{Tiberius}
\label{sec:tiberius}

We used the {\tt Tiberius} pipeline, which builds upon the LRG-BEASTS spectral reduction and analysis pipelines introduced in \citet{Kirk2018,Kirk2019,Kirk2021}, to provide an independent reduction of the data to ensure consistency and robust conclusions. We began our data reduction with our custom group-level background subtracted products from {\tt Tiberius} Stage 1. For each detector, we created bad-pixel masks by manually identifying hot pixels within the data. We then included 5$\sigma$ outlier pixels in the mask, which were identified via running medians operating along the pixel rows. Custom saturated pixel masks were created by flagging a column as saturated if any pixel within that column was saturated for any integration. Prior to identifying the spectral trace, we interpolated each column of the detectors onto a grid 10$\times$ finer than the initial spatial resolution. This step reduces the noise in the extracted data by improving the extraction of flux at the sub-pixel level, particularly where the edges of the photometric aperture bisect a pixel. We also interpolated over the bad pixels using their nearest neighboring pixels in $x$ and $y$. 

We traced the spectra by fitting Gaussians at each column and used a median filter, calculated with a moving box with a width of five pixels in the finely interpolated data, to smooth the measured centers of the trace. We fitted these smoothed centers with a 4th-order polynomial, removed points that deviated from the median by 5$\sigma$, and re-fitted with a 4th-order polynomial. To remove residual background flux not captured by the group-level destriping, we first fit a linear polynomial along each column. We then masked the stellar spectrum, where the mask is defined by an aperture with a width of ten pixels centered on the trace. We also masked an additional four pixels on either side of the aperture so that the background was not fitting the wings of the stellar PSF, and we clipped any pixels within the background that deviated by more than three standard deviations from the mean for that particular column and frame. After removing the background in each column, the stellar spectra were then extracted by summing within the 10-pixel wide aperture and correcting for pixel oversampling caused by the interpolation onto a finer grid, as described above. The uncertainties in the stellar spectra were calculated from the photon noise and read noise.

With the broadband light curves (NRS1: 2.86--3.69\,$\mu$m, NRS2: 3.82--5.06\,$\mu$m) generated from the \texttt{Tiberius} stellar spectra, we fitted for $R_{p}/R_{\star}$, $i$, $T_{0}$, $a/R_{\star}$, the quadratic limb darkening coefficient $u_{1}$, and the coefficients of the systematics model parameters ($x$ and $y$ pixel shifts, FWHM, and background), with $P$, $e$, and fixed to the values given in Table \ref{tab:sys_params} and $u_2$ fixed. We used uniform priors for all the fitted parameters. Our analytic transit light-curve model was generated with {\tt batman} \citep{Kreidberg2015}. We fit our broadband light curves with a transit+systematics model using the Levenberg Marquardt algorithm and assuming a linear polynomial function for each systematic being fit. For the 60 pixel spectroscopic light curves, we held $a/R_{\star}$, $i$, and $T_{0}$ fixed to the best-fit values from the broadband light curve fits (Table \ref{tab:wlc_fits}). We then fit the spectroscopic light curves, again using the Levenberg-Marquadt algorithm with linear polynomial functions for the same systematics detrending parameters detailed above. We used wide uniform priors for all fitted parameters. The {\tt Tiberius} fitted white light curves and residuals are shown in the middle panel of Figure \ref{fig:wlc}, and the best-fit white light curve parameters are shown in Table \ref{tab:wlc_fits}. 

\begin{figure}
\begin{centering}
\includegraphics[trim=1cm 0 0 0,width=0.45\textwidth]{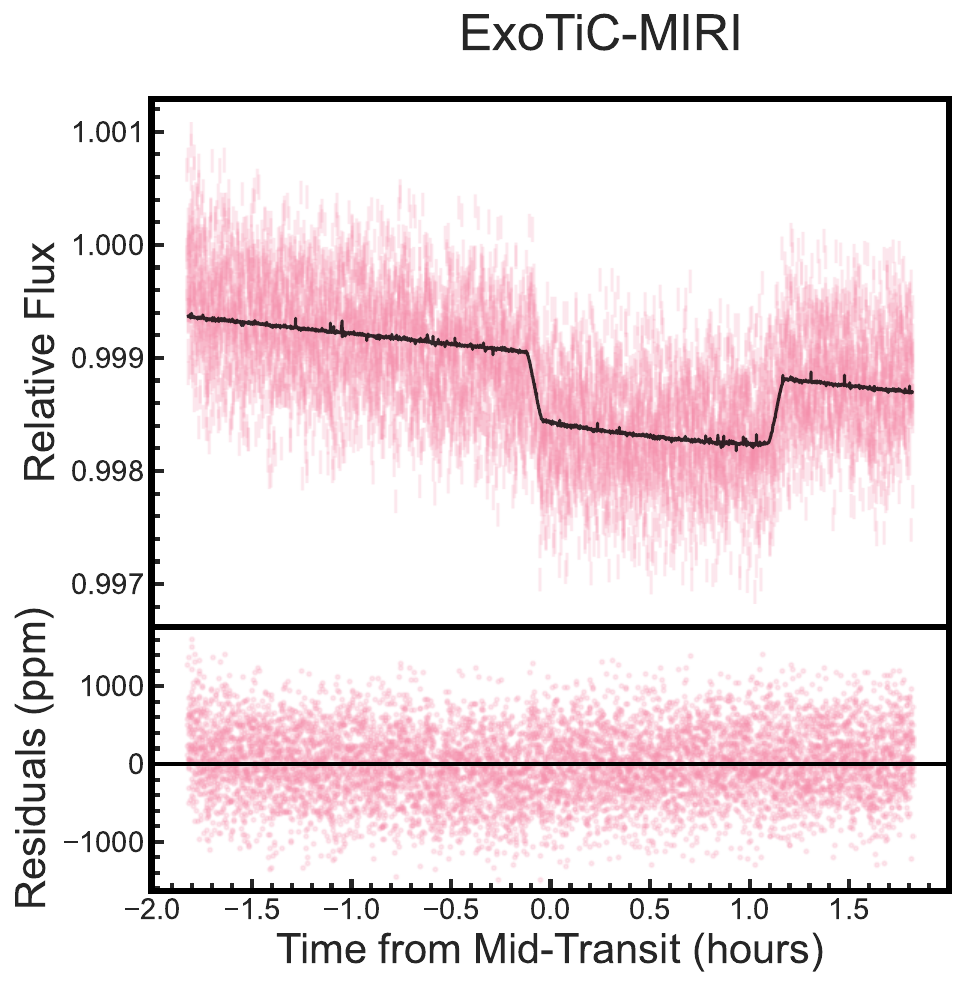}
\caption{ The MIRI/LRS white light curve for L168-9\,b (points) compared to the best-fitting transit+systematics model (black). The residuals are shown in the bottom panel.} 
\label{fig:miri_wlc}   
\end{centering}
\end{figure} 

\begin{figure*}
\begin{centering}
\includegraphics[width=\textwidth]{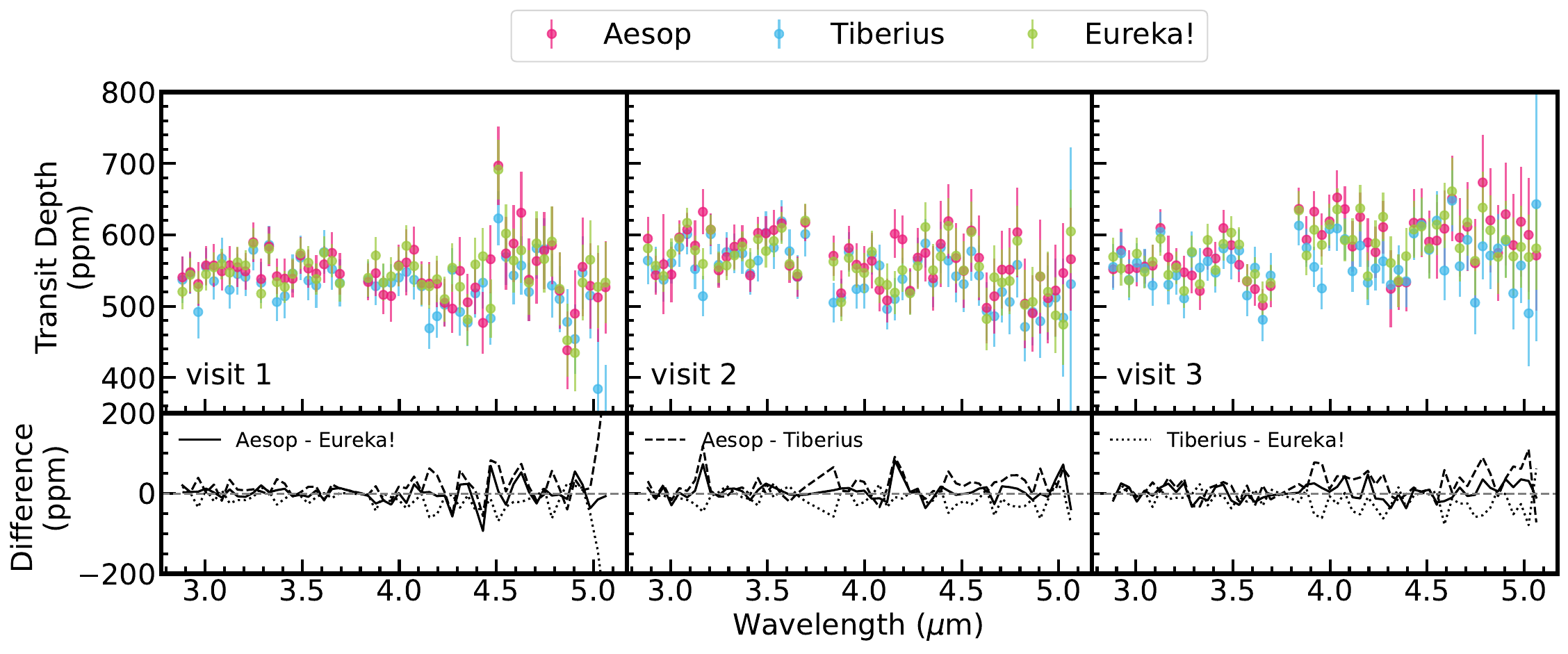}
\includegraphics[width=\textwidth]{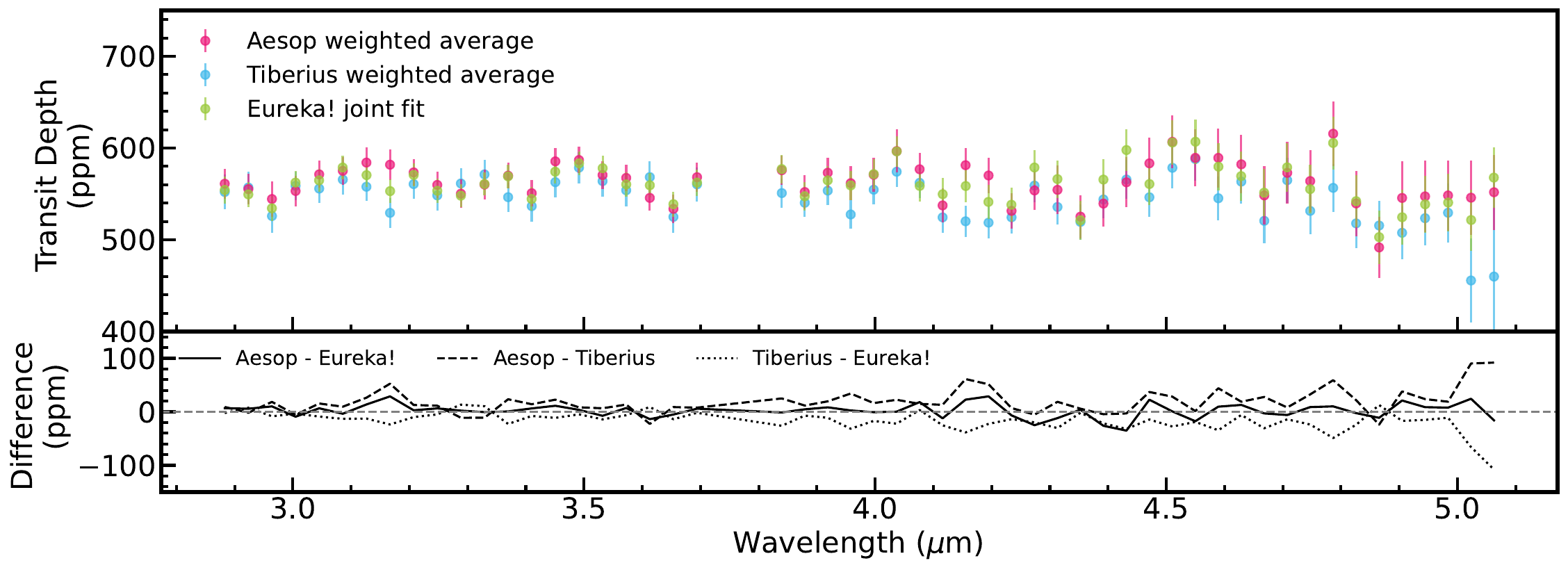}
\caption{\textit{Top}: Comparison of the {\tt Aesop} (pink), {\tt Tiberius} (blue), and {\tt Eureka!} (green) reductions for visit 1 (left), visit 2 (middle), and visit 3 (left). \textit{Bottom}: the combined transmission spectrum for all three visits from the {\tt Aesop} and {\tt Tiberius} weighted averages and the {\tt Eureka!} joint fit.} 
\label{fig:reduction_comp}   
\end{centering}
\end{figure*} 

\begin{figure}
\begin{centering}
\includegraphics[trim=1cm 0 0 0,width=0.45\textwidth]{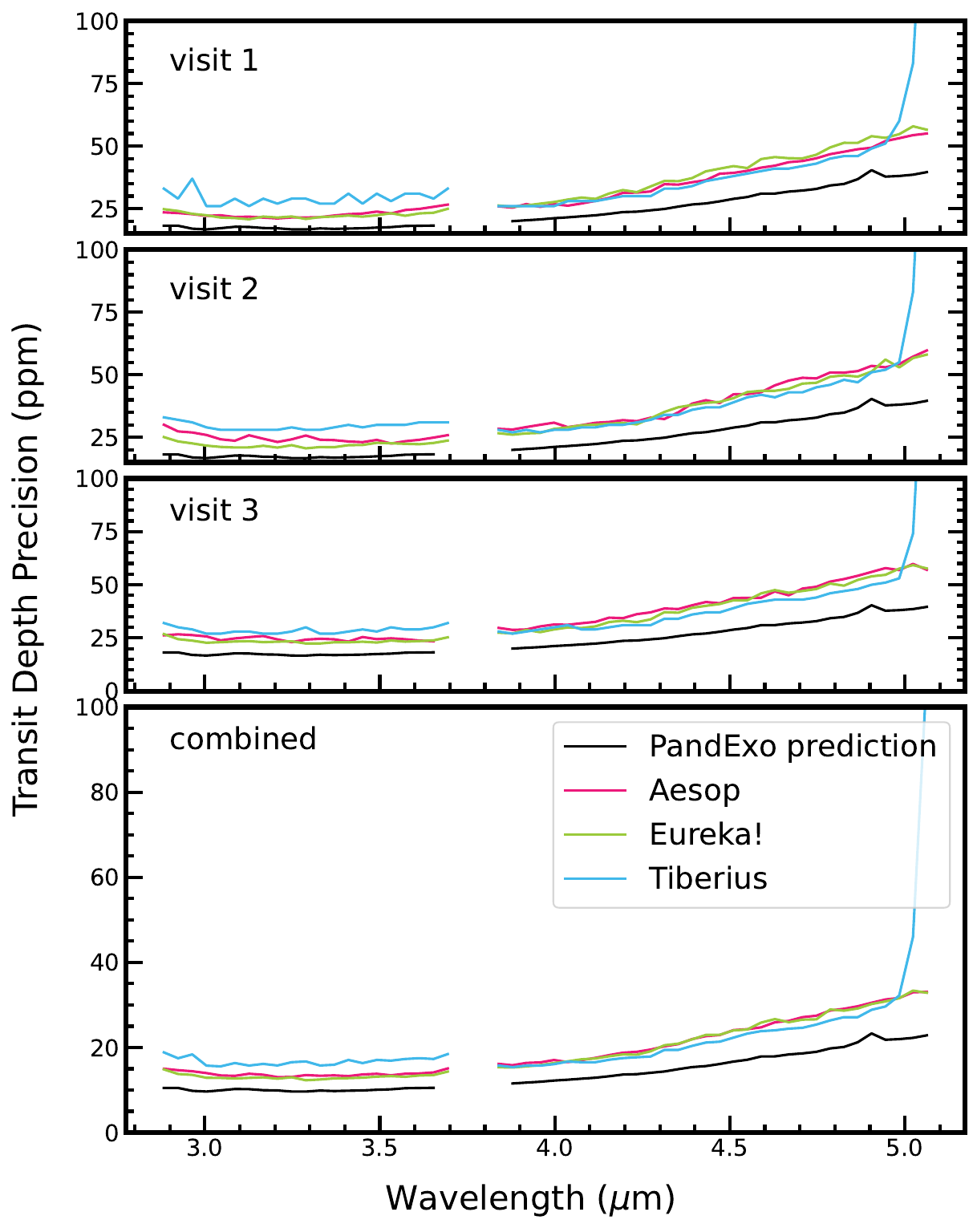}
\caption{The transit depth precisions achieved by each independent reduction compared to the predicted values from {\tt PandExo} simulations for each individual visit and for the combined spectra from each reduction.} 
\label{fig:precisions}   
\end{centering}
\end{figure} 

\begin{figure*}
\begin{centering}
\includegraphics[trim=1cm 0 0 0,width=0.96\textwidth]{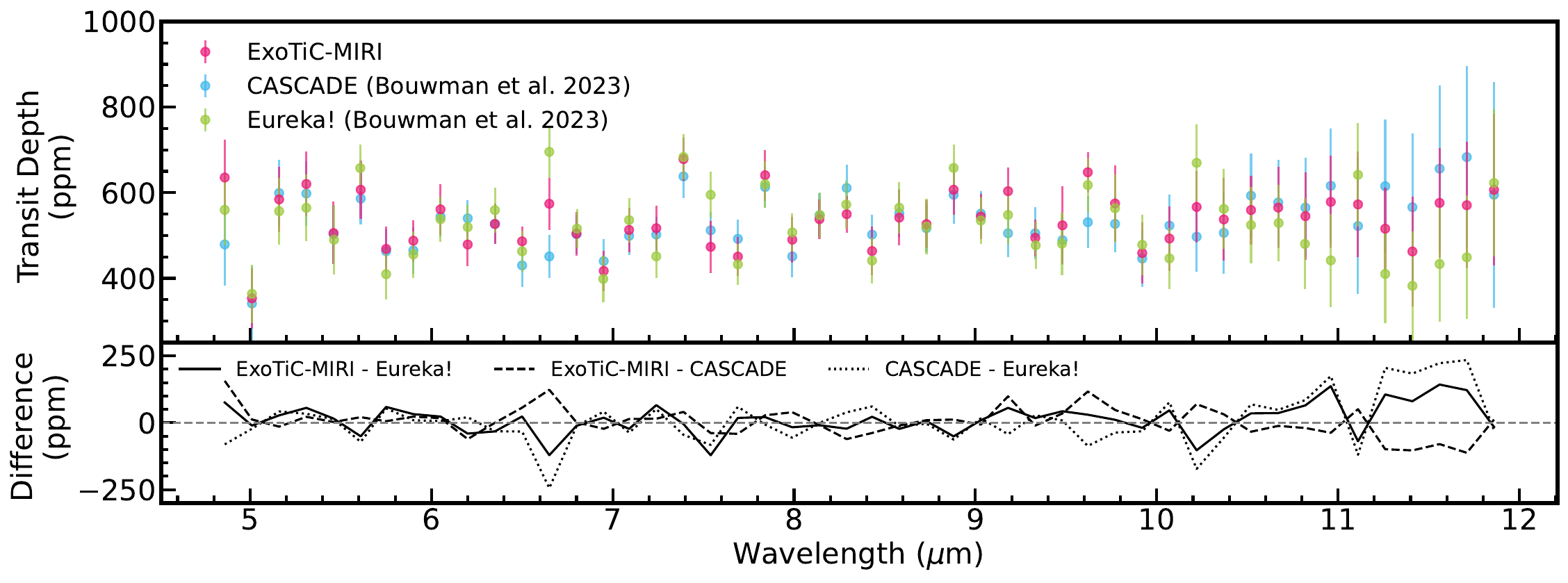}
\caption{\textit{Top:} Comparison of the {\tt ExoTiC-MIRI} (pink) reduction of the MIRI/LRS commissioning data compared to the {\tt CASCADE} (blue) and {\tt Eureka!} (green) reductions from \citet{Bouwman23}.} 
\label{fig:miri_comp}   
\end{centering}
\end{figure*}

\subsubsection{Eureka!}
\label{sec:eureka} 

We also used \texttt{Eureka!}\footnote{\url{https://github.com/kevin218/Eureka}}\,(version 0.10; \citealt{Bell2022}) to provide an additional independent reduction of our observations, treating the data from NRS1 and NRS2 separately throughout our analysis. For Stages 1 and 2 (which follow the standard steps of the first two stages of the \texttt{jwst} pipeline version 1.11.4; \citealt{Bushouse2022}), we used the default \texttt{Eureka!}\,parameters, a jump detection threshold of 15, and the custom 1/$f$ noise correction method from {\tt ExoTiC-JEDI}\footnote{\url{https://github.com/Exo-TiC/ExoTiC-JEDI}} (which leads to fewer temporal outliers). 

We then ran \texttt{Eureka!}\,Stage 3 to extract the time-series stellar spectra and generate white and spectroscopic light curves for each visit. We tested combinations of extraction apertures of 4-8 pixels from the center of the flattened spectral trace, background apertures of 8-11 pixels, and sigma thresholds for optimal extraction outlier rejection of 10 and 60, and two different background subtraction methods (an additional column-by-column mean subtraction and a full frame median subtraction). For each detector and each visit, we selected the combination of reduction parameters that minimizes the RMS scatter in the resulting white light curves. For NRS1, we used aperture half-widths of 4, 6, and 4 and background apertures of 8, 9, and 10, for visits 1, 2, and 3, respectively. For NRS2, we used aperture half-widths of 4, 5, and 4 for visits 1, 2, and 3, respectively and background apertures of 8 for all visits. We used an additional column-by-column background subtraction for NRS1 visits 1 and 2, and a full frame background subtraction for NRS1 visit 3 as well as all visits of NRS2. We selected a 10$\sigma$ threshold for the optimal extraction of NRS1 for visits 2 and 3 and both detectors in visit 1. We also selected a 60$\sigma$ threshold for the optimal extraction of visits 2 and 3 for NRS2.

From our white and 60 pixel spectroscopic light curves, we iteratively trimmed 3$\sigma$ outliers three times from a 50 point rolling median in time. We then used \texttt{emcee} \citep{Foreman-Mackey2013} to fit a combined transit+systematics model to the light curves, fitting for $i$, $a/R_{\star}$, $T_{0}$, and $R_{p}/R_{\star}$. The transit model was generated using \texttt{batman} \citep{Kreidberg2015}, and we fixed the quadratic limb-darkening coefficients to the theoretical values computed with {\tt ExoTiC-LD} \citep{ExoTiC-LD} using the stellar parameters given in Table~\ref{tab:sys_params} and Set One of the MPS-ATLAS models \citep{Kostogryz2022,Kostogryz2023}. Our instrumental noise model, \textit{S}, was of the form
\begin{equation}
S= p_{1} + p_{2}\times T+ p_{3}\times X + p_{4}\times Y , 
\label{eq:1}
\end{equation}
where $p_{N}$ is a parameter fitted for in our instrumental noise model, $T$ is the vector of times, and $X$ and $Y$ are vectors of the positions of the trace from \texttt{Eureka!}\,Stage 3. We trimmed the first 500 points (30 minutes) to remove any initial ramp. We initialized 3$\times$ the number of free parameters as the number of walkers and used a burn-in of 50,000 steps (which is discarded), followed by a production run of 50,000 steps to ensure adequate sampling of the posterior. The fitted white light curves and residuals are shown in the bottom panel of Figure \ref{fig:wlc}, and the best-fitting parameters are given in Table \ref{tab:wlc_fits}. 

For the spectroscopic light curve fits, we combined the MCMC chains from the NRS1 and NRS2 fits for $i$, $a/R_{\star}$, and $T_{0}$. We used the resulting chains as strong Gaussian priors for these parameters centered at the median of each combined chain and with a conservative width of three times the standard deviation of the combined chains. We utilized a wide flat prior for $R_{p}/R_{\star}$ for each spectroscopic bin. We also jointly fit the three visits for each detector in the same manner described above, but assuming a common $i$, $a/R_{\star}$, $T_{0}$, and $R_{p}/R_{\star}$ for all three visits. 

\begin{figure}
\begin{centering}
\includegraphics[trim=1cm 0 0 0,width=0.45\textwidth]{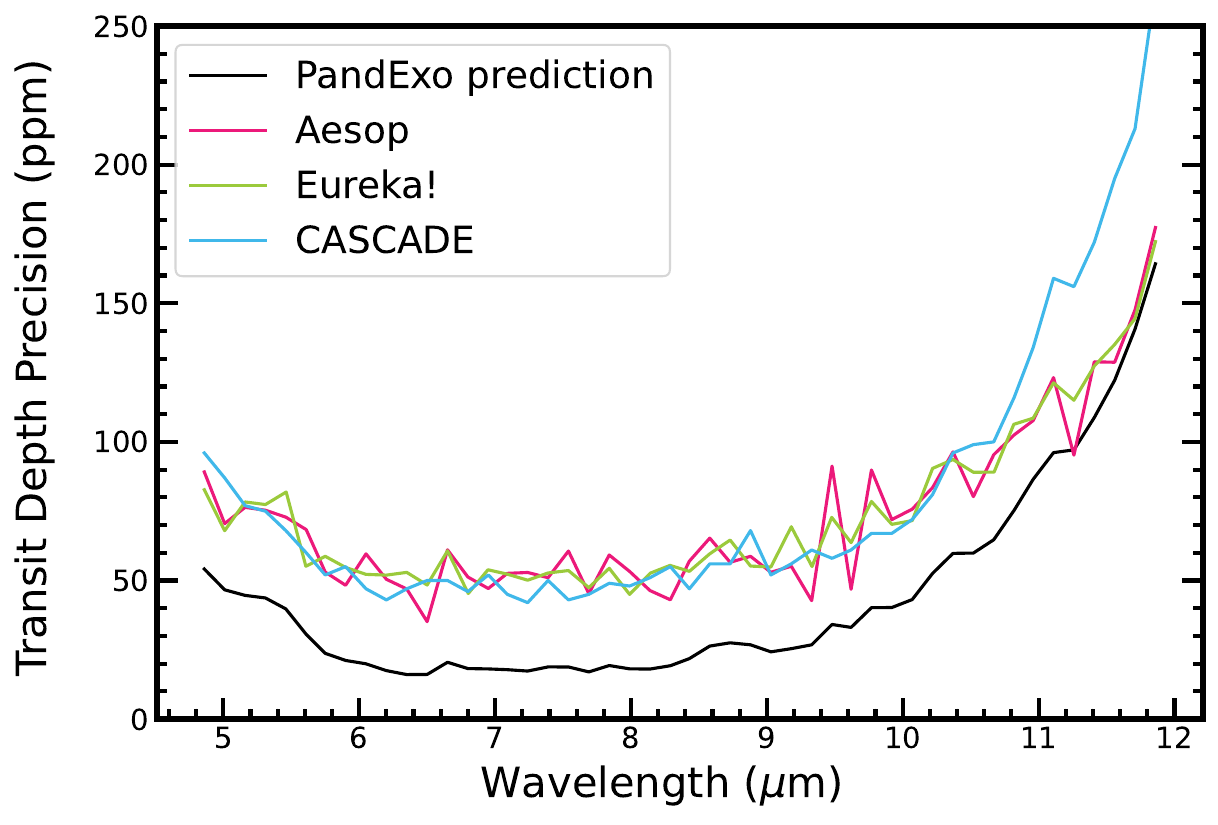}
\caption{ The MIRI transit depth precisions for each independent reduction compared to the predictions from {\tt PandExo}.}
\label{fig:miri_precisions}   
\end{centering}
\end{figure} 

\begin{table*}[]
\centering
\caption{White light curve best-fit values for the {\tt Aesop}, {\tt Tiberius}, {\tt Eureka!}, and {\tt ExoTiC-MIRI} reductions, as well as the {\tt Eureka!} joint fit.}
\label{tab:wlc_fits}
\begin{tabular}{ccc|c|c|c|c}
\hline \hline
\multicolumn{3}{c|}{}                     & $T_{0}$ (MJD) &  $a/R_{\star}$ & $i$ ($\degree$) & $R_{p}/R_{\star}$ \\ \hline \hline
\multicolumn{1}{c|}{}   & \multicolumn{1}{c|}{\multirow{2}{*}{visit 1}} & NRS1 & 60098.46289$\pm5$e${-5}$  & 8.58$\pm$0.80   & 88.32$\pm$1.34 & 0.02381$\pm$0.00017  \\
\multicolumn{1}{c|}{}                                        & \multicolumn{1}{c|}{}                         & NRS2 & 60098.46279$\pm$1e${-4}$   & 7.06$\pm$0.10   & 85.00$\pm$1.01 & 0.02395$\pm$0.00020  \\
\multicolumn{1}{c|}{\multirow{2}{*}{{\tt Aesop}}}   & \multicolumn{1}{c|}{\multirow{2}{*}{visit 2}} & NRS1 & 60113.87973$\pm$4e${-5}$  & 7.86$\pm$0.89   & 89.44$\pm$2.51 & 0.02418$\pm$0.00017  \\
\multicolumn{1}{c|}{}                                        & \multicolumn{1}{c|}{}                         & NRS2 & 60113.87952$\pm$7e${-5}$   & 8.66$\pm$0.17   & 88.32$\pm$2.68 & 0.02316$\pm$0.00019   \\
\multicolumn{1}{c|}{}                                        & \multicolumn{1}{c|}{\multirow{2}{*}{visit 3}} & NRS1 & 60122.28873$\pm$6e${-5}$ & 7.21$\pm$0.82   & 85.31$\pm$2.28 & 0.02429$\pm$0.00016  \\
\multicolumn{1}{c|}{}                                        & \multicolumn{1}{c|}{}                         & NRS2 & 60122.28909$\pm$6e${-5}$   & 7.72$\pm$0.15   & 85.76$\pm$2.08 & 0.02439$\pm$0.00019  \\ \hline
\multicolumn{1}{c|}{}   & \multicolumn{1}{c|}{\multirow{2}{*}{visit 1}} & NRS1 & 60098.46294 $\pm$1e${-4}$  & 8.25$\pm$0.79   & 87.66$\pm$2.20 & 0.02343$\pm$0.00030  \\
\multicolumn{1}{c|}{}                                        & \multicolumn{1}{c|}{}                         & NRS2 & 60098.463000$\pm$2e${-4}$   & 7.11$\pm$0.76   & 85.08$\pm$1.59 & 0.02353$\pm$0.00031  \\
\multicolumn{1}{c|}{\multirow{2}{*}{{\tt Tiberius}}}   & \multicolumn{1}{c|}{\multirow{2}{*}{visit 2}} & NRS1 & 60113.879977$\pm$1e${-4}$  & 7.52$\pm$0.63  & 85.93$\pm$1.34 & 0.02416$\pm$0.00029 \\
\multicolumn{1}{c|}{}                                        & \multicolumn{1}{c|}{}                         & NRS2 & 60113.87968 $\pm$2e${-4}$   & 8.28$\pm$0.91   & 87.73$\pm$2.57 & 0.02371$\pm$0.00018  \\
\multicolumn{1}{c|}{}                                        & \multicolumn{1}{c|}{\multirow{2}{*}{visit 3}} & NRS1 & 60122.289124$\pm$2e${-4}$ & 6.93$\pm$0.56   & 84.72$\pm$1.16 & 0.02363$\pm$0.00029\\
\multicolumn{1}{c|}{}                                        & \multicolumn{1}{c|}{}                         & NRS2 & 60122.289356$\pm$2e${-4}$  &  7.03$\pm$0.64   & 84.91$\pm$1.32 & 0.02491$\pm$0.00033  \\ \hline
\multicolumn{1}{c|}{}   & \multicolumn{1}{c|}{\multirow{2}{*}{visit 1}} & NRS1 & 60098.46286$\pm$5e${-5}$  & 8.71$\pm$0.40   & 89.33$\pm$1.73 & 0.02348$\pm$0.00016  \\
\multicolumn{1}{c|}{}                                        & \multicolumn{1}{c|}{}                         & NRS2 & 60098.46270$\pm$8e${-5}$   & 8.28$\pm$0.58   &  87.44$\pm$2.04 & 0.02344$\pm$0.00020  \\
\multicolumn{1}{c|}{\multirow{2}{*}{{\tt Eureka!}}}   & \multicolumn{1}{c|}{\multirow{2}{*}{visit 2}} & NRS1 & 60113.87982$\pm$6e${-5}$  & 8.58$\pm$0.46   & 88.74$\pm$1.92 & 0.02409$\pm$0.00018  \\
\multicolumn{1}{c|}{}                                        & \multicolumn{1}{c|}{}                         & NRS2 & 60113.87960$\pm$6e${-5}$   & 8.76$\pm$0.34   & 89.40$\pm$1.58 & 0.02346$\pm$0.00016  \\
\multicolumn{1}{c|}{}                                        & \multicolumn{1}{c|}{\multirow{2}{*}{visit 3}} & NRS1 & 60122.28884$\pm$6e${-5}$ & 8.20$\pm$0.60   & 87.26$\pm$2.02 & 0.02369$\pm$0.00022 \\
\multicolumn{1}{c|}{}                                        & \multicolumn{1}{c|}{}                         & NRS2 & 60122.28894$\pm$7e${-5}$   & 8.01$\pm$0.60   & 86.86$\pm$1.97 & 0.02455$\pm$0.00020  \\ \hline
\multicolumn{1}{c|}{}                                        & \multicolumn{1}{c|}{\multirow{6}{*}{}}   & \multirow{3}{*}{NRS1} & 60098.46292$\pm$4e${-5}$                     & \multirow{3}{*}{$8.39\pm0.47$} & \multirow{3}{*}{$87.74\pm1.61$} & \multirow{3}{*}{0.02379$\pm$0.00015} \\
\multicolumn{1}{c|}{}                                        & \multicolumn{1}{c|}{}                         &                       & 60113.87972$\pm$4e${-5}$                    &                       &                    &     \\
 \multicolumn{1}{c|}{{\tt Eureka!} joint fit}                                        & \multicolumn{1}{c|}{all visits}                         & \multirow{3}{*}{NRS2} & 60122.28888$\pm$4e${-5}$                  &                        &       &   \\          
\multicolumn{1}{c|}{}                                        & \multicolumn{1}{c|}{}                         &                       & 60098.46284$\pm$4e${-5}$         & \multirow{3}{*}{8.72$\pm$0.30} & \multirow{3}{*}{89.43$\pm$1.55} & \multirow{3}{*}{0.02358$\pm$0.00028} \\\multicolumn{1}{c|}{}                                        & \multicolumn{1}{c|}{}                         &                       & 60113.87964$\pm$4e${-5}$                   &                        &         \\
\multicolumn{1}{c|}{}                                        & \multicolumn{1}{c|}{}                         &                       & 60122.28880$\pm$4e${-5}$                   &                        &         \\
\hline 
\multicolumn{1}{c|}{\multirow{1}{*}{{\tt ExoTiC-MIRI}}}   & \multicolumn{1}{c|}{\multirow{1}{*}{visit 1}} & LRS & 59082.85642$\pm$8e${-5}$  & 8.24$\pm$0.64   & 88.1$\pm$1.25 & 0.02333$\pm$0.00028  \\ \hline                

\end{tabular}
\end{table*}

\subsection{MIRI LRS}

\subsubsection{ExoTiC-MIRI}

We reduced the commissioning observations of L\,168-9\,b using the {\tt ExoTiC-MIRI}\footnote{\url{https://exotic-miri.readthedocs.io/en/latest/}} pipeline \citep{ExoTiC-MIRI}. We began with the uncalibrated files from the {\tt jwst} pipeline (version 1.8.2, context map 1078; \citealt{Bushouse2022}). We then processed the data using the standard Stage 1 steps for time-series observations, as well as custom steps unique to the {\tt ExoTiC-MIRI} pipeline. We started with the data quality initialization step, followed by corrections for saturation, linearity, and dark current. We applied the {\tt custom}$\textunderscore{{\tt drop}}\textunderscore{{\tt groups}}$ step to drop the first five groups before the linearity correction, since these groups may adversely affect the group-level ramps due to detector effects such as reset switch charge decay as well as the last frame effect pulling down the final group \citep{Ressler15,Wright23}. For the jump step, we set the detection threshold to 15 to prevent spurious cosmic ray flagging. We then proceeded with the standard ramp fitting and gain scale correction steps, where we set the gain value to 3.1 electrons per data number \citep{Bell23}. 

In Stage 2, we processed the {\tt rateints} files with the standard flat fielding step, and applied custom cleaning and background subtraction routines. The custom cleaning step replaced known bad pixels and removed unidentified outliers from the data quality arrays by estimating a spatial profile from polynomial fits to the detector columns and iteratively replacing pixels by the profile value if they were flagged as {\tt DO}$\textunderscore{{\tt NOT}}\textunderscore{{\tt USE}}$ or were more than 4$\sigma$ discrepant. In the custom background step, we subtracted a row-by-row background using the median value from columns 8 to 17 and 56 to 72. We then performed the spectral extraction using a fixed-width box aperture centered on column 36 and extended four pixels to the left and right to obtain the time-series stellar spectra. 

We extracted the broadband MIRI/LRS light curve by summing up the stellar flux in the time-series between 4.785 and 11.935 $\mu$m. For the spectroscopic light curves, we generated 48 wavelength bins $\sim$0.15\,$\mu$m wide following the binning scheme of \citet{Bouwman23}. We iteratively removed points in the white and spectroscopic light curves that were 4$\sigma$ outliers in the residuals, and trimmed the first 1126 integrations ($\sim$30  minutes) in the time-series to remove the detector settling ramp. We then fit each light curve with a transit model generated using {\tt batman} multiplied by a systematics model of the form:

\begin{equation}
    S(\lambda) = c_{0} + (c_{1} \times x) + (c_{2} \times y) + (c_{3} \times t)
\end{equation}

\noindent where  $x$ and $y$ are the $x$- and $y$- pixel positional shifts on the detector, $t$ is time, and  the coefficients $c_{0}$, $c_{1}$, and $c_{2}$ are constants to be fit.

We then fit the broadband LRS white light curve using a least-squares minimizer. We fixed $P$ to the \citet{Astudillo-Defru20} value, setting $e$ to zero and $\omega$ to ninety, and fit for $T_{0}$, $a/R_{\star}$, $i$, and $R_{p}/R_{\star}$ using wide uniform priors. We adopted the optimized parameters and their standard deviations as the best-fit white light curve values, which are given in Table \ref{tab:wlc_fits}. For the spectroscopic light curve fits, we fixed  $T_{0}$, $a/R_{\star}$, and $i$ to best-fit values from the white light curves, and fit for $R_{p}/R_{\star}$. We fixed the quadratic limb darkening coefficients to the theoretical values calculated using {\tt ExoTiC-LD} \citep{ExoTiC-LD} and Set One of the MPS-ATLAS stellar models \citep{Kostogryz2022,Kostogryz2023} and the stellar $\rm T_{eff}$, log(g), and [Fe/H]$_{\star}$ given in Table \ref{tab:sys_params}. The fitted {\tt ExoTiC-MIRI} white light curve and residuals is shown in Figure \ref{fig:miri_wlc}. 


\section{Interpretation of L\,168-9\,\MakeLowercase{b's} Transmission Spectrum} 
\label{sec:results}

The 3-5\,$\mu$m transmission spectrum using a 60 pixel binning scheme for L\,168-9\,b from NIRSpec/G395H for all three visits and all three reductions are shown in the top panels of Figure \ref{fig:reduction_comp}. Each visit is generally consistent across the three independent reductions, with a median difference in transit depth of 42\,ppm across visits for the {\tt Aesop} reduction (compared to the median transit depth uncertainty of 35\,ppm), 39\,ppm for the  {\tt Tiberius} reduction (median transit depth uncertainty of 54\,ppm), and 41\,ppm for the {\tt Eureka!} reduction (median transit depth uncertainty of 29\,ppm).  We do not apply any offsets between the NRS1 and NRS2 detectors, although we note a median offset of 49\,ppm between NRS1 and NRS2 in visit 3 across all three reductions.  Across visits, the three reductions are also similar, with a median difference between the {\tt Aesop} and {\tt Eureka!} reductions of 18\,ppm, 26\,ppm, and 19\,ppm for visits 1, 2, and 3, respectively. The {\tt Aesop} and {\tt Tiberius} reductions have a median difference of 21\,ppm, 22\,ppm, and 26\,ppm for the three visits, whereas the median transit depth difference between the {\tt Tiberius} and {\tt Eureka!} reductions is 23\,ppm, 34\,ppm, and 26\,ppm.  

The combined spectra for each reduction (the weighted average of the three visits for the {\tt Aesop} and {\tt Tiberius} reductions, as well as the {\tt Eureka!} joint fit) are shown in the bottom panel of Figure \ref{fig:reduction_comp}. No offsets have been applied to the combined spectra. The median transit depth uncertainties for the  {\tt Aesop} weighted average, {\tt Tiberius} weighted average, and {\tt Eureka!} joint fit are 20\,ppm, 28\,ppm, and 18\,ppm, respectively. Taken together, the reductions are consistent within the uncertainties with a median transit depth difference of 10\,ppm between the {\tt Aesop} and {\tt Eureka!} reductions, 12\,ppm between {\tt Aesop} and {\tt Tiberius}, and 15\,ppm between the {\tt Tiberius} and {\tt Eureka!} reductions. We also compare the transit depth precisions from each independent reduction to the predicted precisions from {\tt PandExo} simulations in Figure \ref{fig:precisions}. Despite the fact that the three reductions are comparable, none of them acheive the precisions predicted from the {\tt PandExo} simulations across all NRS1 and NRS2 wavelengths. On average, the {\tt Aesop}, {\tt Tiberius}, and {\tt Eureka} reductions are 1.4$\times$, 2.2$\times$, and 1.3$\times$ the {\tt PandExo} value for each visit and 1.3$\times$, 2.1$\times$, and 1.2$\times$ the {\tt PandExo} value for the combined spectra.    

The precisions achieved by our {\tt ExoTiC-MIRI} reduction are comparable to the {\tt Eureka!} and {\tt CASCADE} reductions reported in \citet{Bouwman23}, as shown in Figure \ref{fig:miri_precisions}. The mid-infrared transmission spectra extracted using {\tt ExoTiC-MIRI}, {\tt Eureka!}, and {\tt CASCADE} are 1.7$\times$, 1.7$\times$, and 1.8$\times$ the predicted value from {\tt PandExo} simulations, respectively.

None of the individual NIRSpec/G395H visits across the three independent reductions show evidence of gaseous molecular absorption features. Our spectrum from the reduction of the MIRI/LRS commissioning observations is shown in Figure \ref{fig:miri_comp}, which also appears to be featureless albeit with larger uncertainties. We consider the near-infrared NIRSpec/G395H spectrum as well as the combined near- to mid-infrared NIRSpec/G395H+MIRI/LRS transmission spectrum in our interpretation of this planet's atmospheric properties (\S \ref{sec:1D_models}).

\subsection{Forward Model Fits}
\label{sec:1D_models}

\begin{figure*}
\begin{centering}
\includegraphics[width=\textwidth]{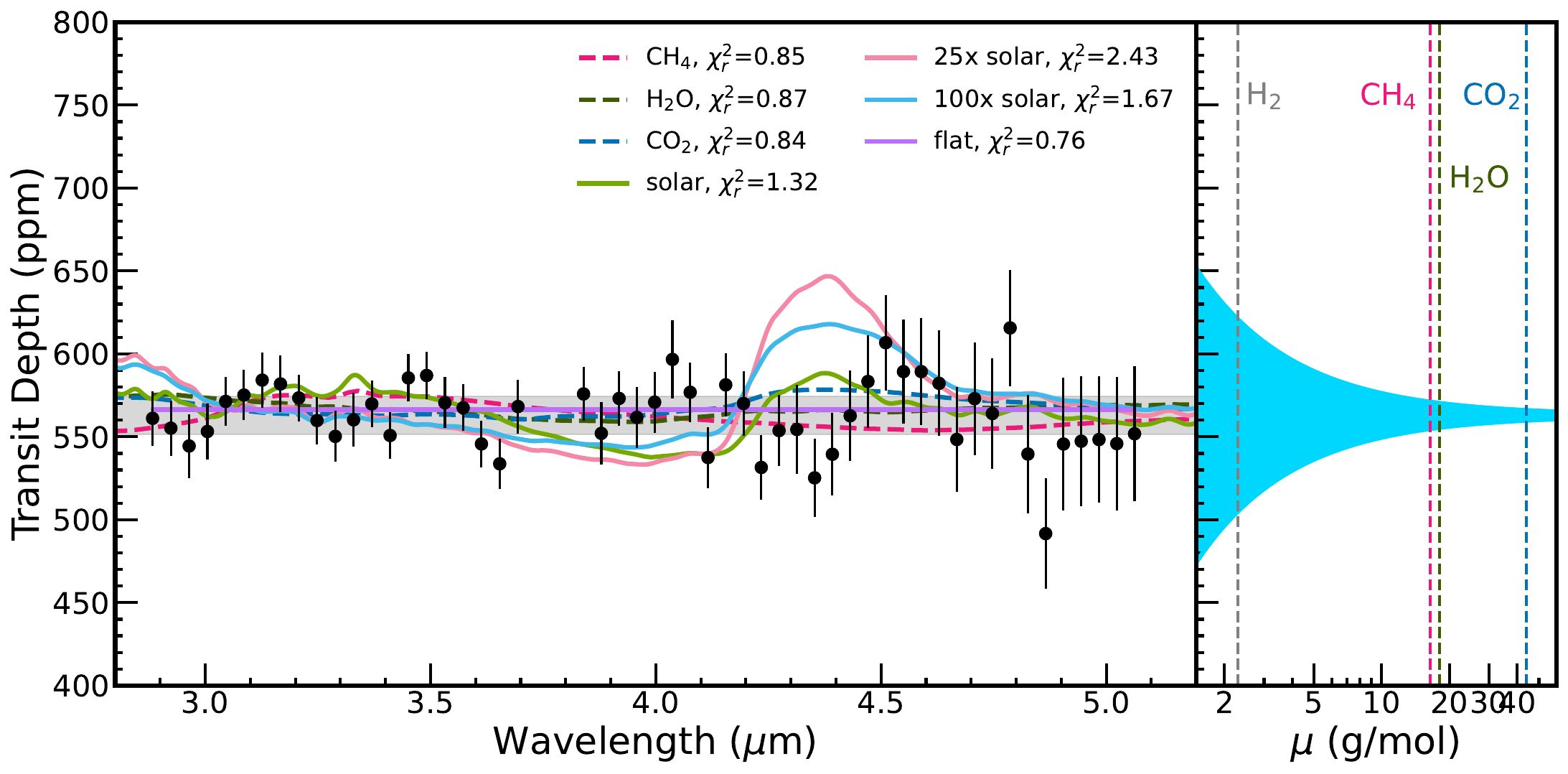}
\includegraphics[trim=0.5cm 0 0 0,width=\textwidth]{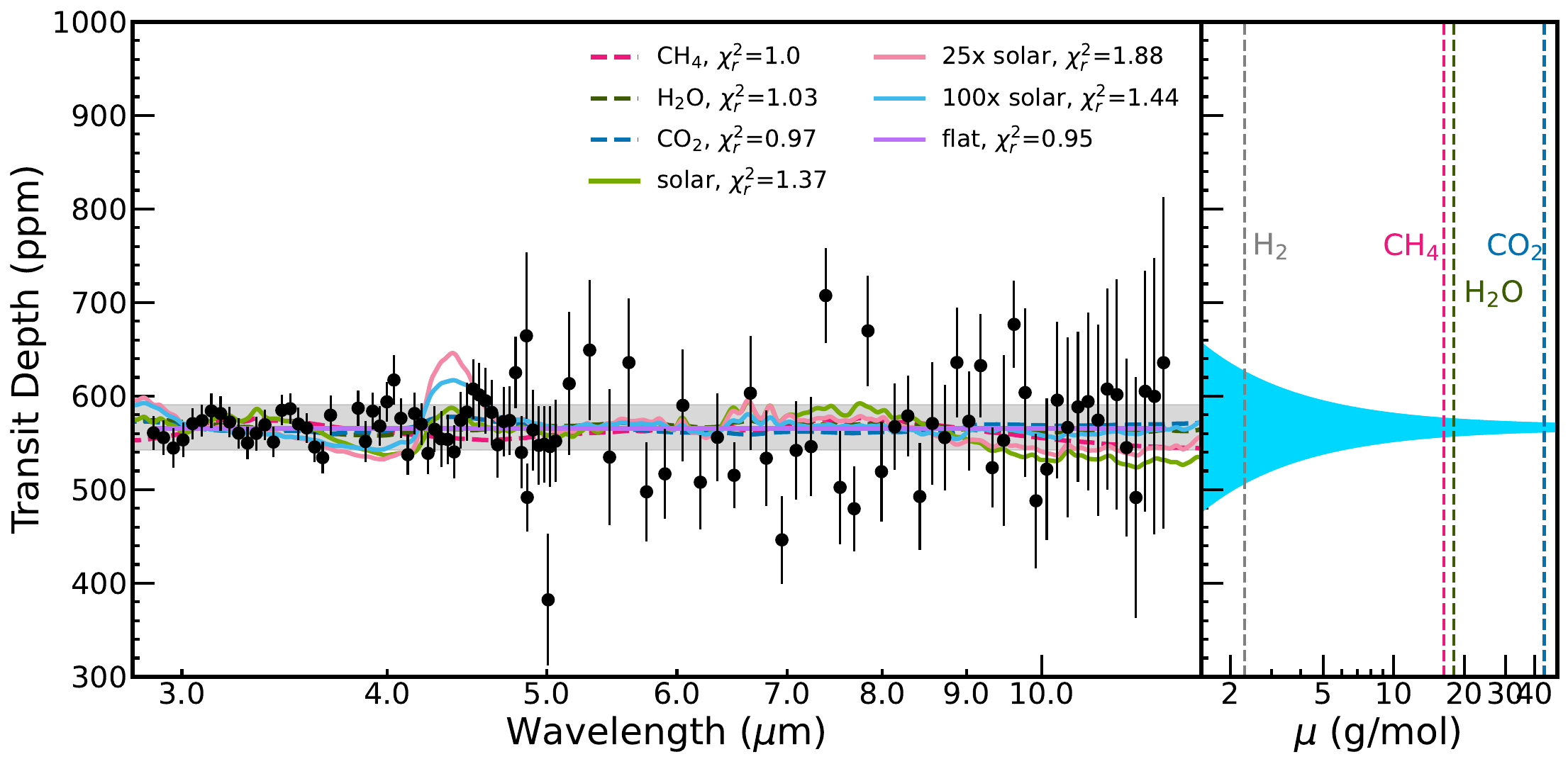}
\caption{\textit{Top}: Left: The {\tt Aesop} weighted average NIRSpec/G395H transmission spectrum (black points) of L\,168-9\,b compared to 1D isothermal atmospheric models assuming single gas compositions, with methane (dark pink, dashed), water (dark green, dashed), and carbon dioxide (dark blue, dashed), as well as solar composition (light green), 25$\times$ solar metallicity (light pink), and 100$\times$ solar (light blue) models extracted from the model grid generated using the analytical TP profile from \citet{Guillot2010}. A flat line model is shown in purple. The best-fitting ($\chi_{r}$=0.76) atmospheric model is the flat line model, consistent with either a high-altitude gray cloud deck, atmospheres with mean molecular weights greater than 4 g\,mol$^{-1}$, or no atmosphere. The light gray shaded region corresponds to the 1$\sigma$ scatter in the data. Right: Typical size of spectral features \citep[$\pm$5 atmospheric scale heights;][]{Seager2000} as a function of mean molecular weight (MMW) compared to the MMW of molecular hydrogen (2.3 g/mol; gray), methane (16.0 g/mol; pink), water (18.02 g/mol; green), and carbon dioxide (44.01 g/mol; blue). The transit depth uncertainty per spectral bin as well as the scatter in the transmission spectrum far exceeds the scale height of any possible high mean molecular weight atmosphere for this planet. \textit{Bottom}: The same as the top panel, except for the NIRSpec/G395H ({\tt Aesop} weighted average) + MIRI/LRS transmission spectrum. A flat line best fits the near- to mid-infrared transmission spectrum ($\chi_{r}^{2}$=0.95).} 
\label{fig:aesop_avg_tr_spec}   
\end{centering}
\end{figure*} 

\begin{figure*}
\begin{centering}
\includegraphics[width=\textwidth]{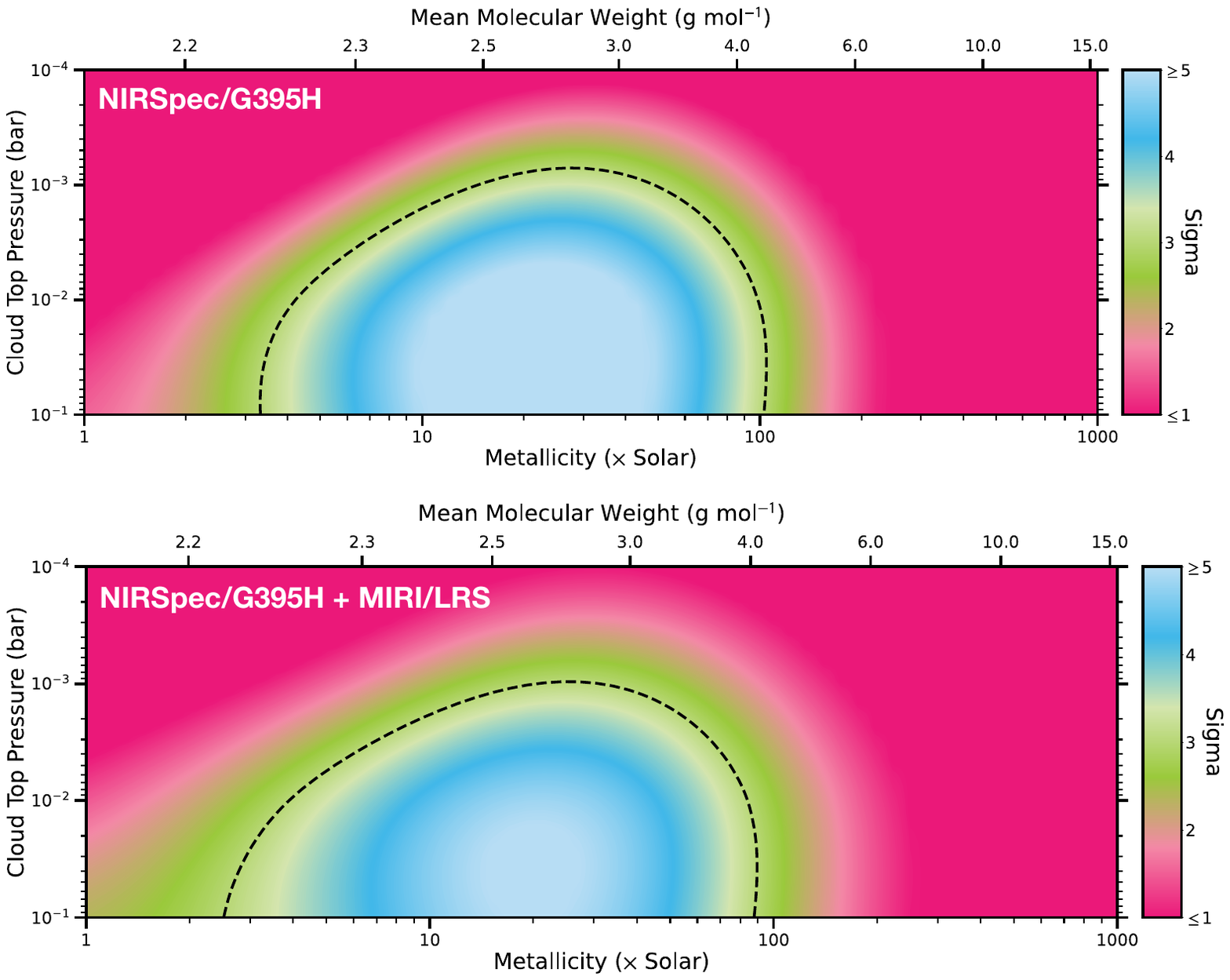}
\caption{\textit{Top}: Cloud top pressure versus atmospheric metallicity for the {\tt Aesop} weighted average NIRSpec/G395H transmission spectrum of L\,168-9b, where the colored contours correspond to the significance level (in $\sigma$) to which we can rule out our forward models. The dashed black curve indicates the threshold beyond which we can rule out regions of this parameter space to $>$3$\sigma$ confidence. \textit{Bottom}: Same as the top panel, but for the NIRSpec/G395H+MIRI/LRS transmission spectrum of L\,168-9b.} 
\label{fig:sigma}   
\end{centering}
\end{figure*} 

We produced a small grid of atmospheric models to interpret the measured transmission spectrum of L\,168-9\,b and quantify the part of parameter space ruled out by our data, as in previous papers in this program \citep{Alderson2024,Wallack2024, Scarsdale_submitted}. Briefly, we consider the 5-parameter double-gray analytical temperature-pressure profile of \citet{Guillot2010}, with which we generated atmospheric abundance profiles using the chemical equilibrium grid from \citet{Line2013}\footnote{\url{https://github.com/mrline/CHIMERA}} calculated using NASA CEA \citep{gordon1994computer}. We included abundances for \ce{H2O}, \ce{CH4}, CO, \ce{CO2}, \ce{NH3}, \ce{N2}, HCN, \ce{H2S}, \ce{PH3}, \ce{C2H2}, \ce{C2H6}, Na, K, TiO, VO, Fe, H, \ce{H2}, and He. We varied the atmospheric metallicity from 0.01 to 1000$\times$ solar and fixed the C/O to solar \citep[0.55;][]{Asplund2009}. A cloud deck with a total optical depth of 10 is included in the model to represent the opaque pressure level, which could be due to gray aerosols or the solid surface of the planet that prevent us from probing deeper pressures. The base of the cloud is set to 10 bars, with the cloud top pressure varying from 1 to 10$^{-4}$ bars. We generated spectra from these atmospheric models using the open-source atmospheric modeling code {\tt PICASO}\footnote{\url{https://github.com/natashabatalha/picaso}} \citep{batalha2019,Mukherjee_2023}, with opacities from \citet{natasha_batalha_2020_3759675} for a subset of the above list of species (Na, K, \ce{CH4}, CO, \ce{CO2}, \ce{H2O}, \ce{H2S}, \ce{NH3}, \ce{PH3}, TiO, and VO). The spectral resolution of the opacities are sufficient for fitting to the R$\sim$100 final observed transmission spectra. In addition to the model grid, we also generated several ``end-member'' atmospheric models consisting of single gases (\ce{CH4}, \ce{H2O}, and \ce{CO2}), for which we assumed an isothermal temperature profile set to the equilibrium temperature of the planet assuming zero albedo and full heat redistribution ($\rm T_{eq}$=981\,K). Finally, we considered a flat line model to represent a high altitude gray cloud or the lack of an atmosphere. 

\begin{table*}[]
\centering
\caption{Reduced chi-squared $\chi_{r}^{2}$ values from the forward model fits described in Section \ref{sec:1D_models}. }
\label{tab:model_fits}
\begin{tabular}{cc|c|c|c|c|c|c|c|c}
\hline \hline
\multicolumn{2}{c|}{}    & $\ce{CH4}$  &  $\ce{CO2}$  & $\ce{H2O}$  & 1$\times$ solar &  25$\times$ solar & 100$\times$ solar & flat \\ \hline \hline 

\multicolumn{1}{c|}{}   & \multicolumn{1}{c|}{\multirow{1}{*}{{\tt Aesop}}} & 0.54 & 0.62 & 0.58 & 0.66  & 1.21 & 0.93  & 0.57 \\
\multicolumn{1}{c|}{\multirow{1}{*}{G395H visit 1}}   & \multicolumn{1}{c|}{\multirow{1}{*}{{\tt Tiberius}}} & 0.68  & 0.62 & 0.78 & 0.92 & 1.26 & 1.00 & 0.71 \\
\multicolumn{1}{c|}{}                                        & \multicolumn{1}{c|}{\multirow{1}{*}{{\tt Eureka!}}} & 0.82 & 0.86 & 0.83 & 1.16 & 1.98 & 1.56 & 0.76 \\
\hline
\multicolumn{1}{c|}{}   & \multicolumn{1}{c|}{\multirow{1}{*}{{\tt Aesop}}} & 0.83  & 0.89 &  0.89 & 0.86  & 1.22 & 1.01  & 0.88 \\
\multicolumn{1}{c|}{\multirow{1}{*}{G395H visit 2}}   & \multicolumn{1}{c|}{\multirow{1}{*}{{\tt Tiberius}}} & 0.38 & 0.41 & 0.40 & 1.48 & 1.53 & 1.45 & 0.40 \\

\multicolumn{1}{c|}{}                                        & \multicolumn{1}{c|}{\multirow{1}{*}{{\tt Eureka!}}} & 1.17 & 1.13 & 1.18 & 1.07 & 2.02 & 1.78 & 1.07 \\
\hline
\multicolumn{1}{c|}{}   & \multicolumn{1}{c|}{\multirow{1}{*}{{\tt Aesop}}} & 1.34 & 1.13 & 1.24 & 1.73  & 1.78 & 1.45 & 1.14\\
\multicolumn{1}{c|}{\multirow{1}{*}{G395H visit 3}}   & \multicolumn{1}{c|}{\multirow{1}{*}{{\tt Tiberius}}} & 0.86 & 0.88 & 0.87 & 1.26 & 1.23 & 1.03 & 0.87 \\
\multicolumn{1}{c|}{}                                        & \multicolumn{1}{c|}{\multirow{1}{*}{{\tt Eureka!}}} & 1.36 & 1.42 & 1.21 & 1.93 & 2.02 & 1.57 & 0.88 \\
\hline
\multicolumn{1}{c|}{}   & \multicolumn{1}{c|}{\multirow{1}{*}{{\tt Aesop}}} & 0.85 & 0.87 & 0.84 & 1.29  & 2.42 & 1.67 &  0.76 \\
\multicolumn{1}{c|}{\multirow{1}{*}{G395H combined}}   & \multicolumn{1}{c|}{\multirow{1}{*}{{\tt Tiberius}}} & 0.86 & 0.86 & 0.68 & 1.45 & 1.89 & 1.35 & 0.87 \\

\multicolumn{1}{c|}{}                                        & \multicolumn{1}{c|}{\multirow{1}{*}{{\tt Eureka!}}} & 1.18 & 1.11 & 1.14 & 1.73 & 3.5 & 2.41 & 1.08 \\
\hline
 \multicolumn{1}{c|}{\multirow{1}{*}{G395H combined + MIRI}}   & \multicolumn{1}{c|}{\multirow{1}{*}{{\tt Aesop}}} & 1.0 & 1.03 & 0.97 & 1.37   & 1.86 & 1.42  & 0.95 \\

\hline

\end{tabular}
\end{table*}

We fitted the aforementioned models to the near-infrared NIRSpec/G395H (3--5\,$\mu$m) transmission spectrum only, as well as to the full NIRSpec/G395H+MIRI/LRS (3--12\,$\mu$m) spectrum by optimizing the vertical offsets of the model spectra. 
The number of degrees of freedom for each model is constant and computed as $n$ -- $m$, where $n$ is the number of data points ($n$=54 for the G395H-only fits and $n$=102 for the G395H+MIRI fits) and $m$=1 is the number of fitted parameters (vertical offset). For the G395H+MIRI fits, we shifted the MIRI data vertically by 29\,ppm such that the mean transit depth across the MIRI wavelengths matched that of the G395H data. From our fits, we quantified our model selection by computing the $\chi^{2}$ statistic. The results of our model fitting exercises are shown in Table \ref{tab:model_fits}. Henceforth in the paper, we focus on the {\tt Aesop} weighted average spectrum in our interpretation of L\,168-9\,b's atmosphere, since we demonstrated in Section \ref{sec:results} that all of the independent reductions presented here are consistent within 1$\sigma$.

The  {\tt Aesop} weighted average G395H and G395H+MIRI transmission spectra of L\,168-9\,b compared to the single-gas and flat models, as well as several illuminating models from the model grid (1$\times$, 25$\times$, and 100$\times$ solar with opaque pressure levels at 1 bar) are shown in Figure \ref{fig:aesop_avg_tr_spec}. In addition, we computed the typical size of spectral features \citep[$\sim$5 scale heights;][]{Seager2000} as a function of atmospheric mean molecular weight for comparison to the data using Equation 1 from \citet{stevenson2016}. We find that both datasets prefer the single gas models and the flat model over any of the super solar metallicity models, with the solar metallicity model possessing intermediate $\chi_r^2$, likely due to the lack of high amplitude molecular features (e.g., CO$_2$). This finding suggests a high ($>$4 g mol$^{-1}$) atmospheric mean molecular weight, the presence of high altitude aerosols, or the lack of an atmosphere altogether. Among the relatively well-fit models, the $\chi_r^2$ values are all similar, and the lack of any discernible atmospheric features prevents us from discriminating between them. 

\begin{figure*}
\begin{centering}
\includegraphics[trim=2cm 0 0 0, width=1.15\textwidth]{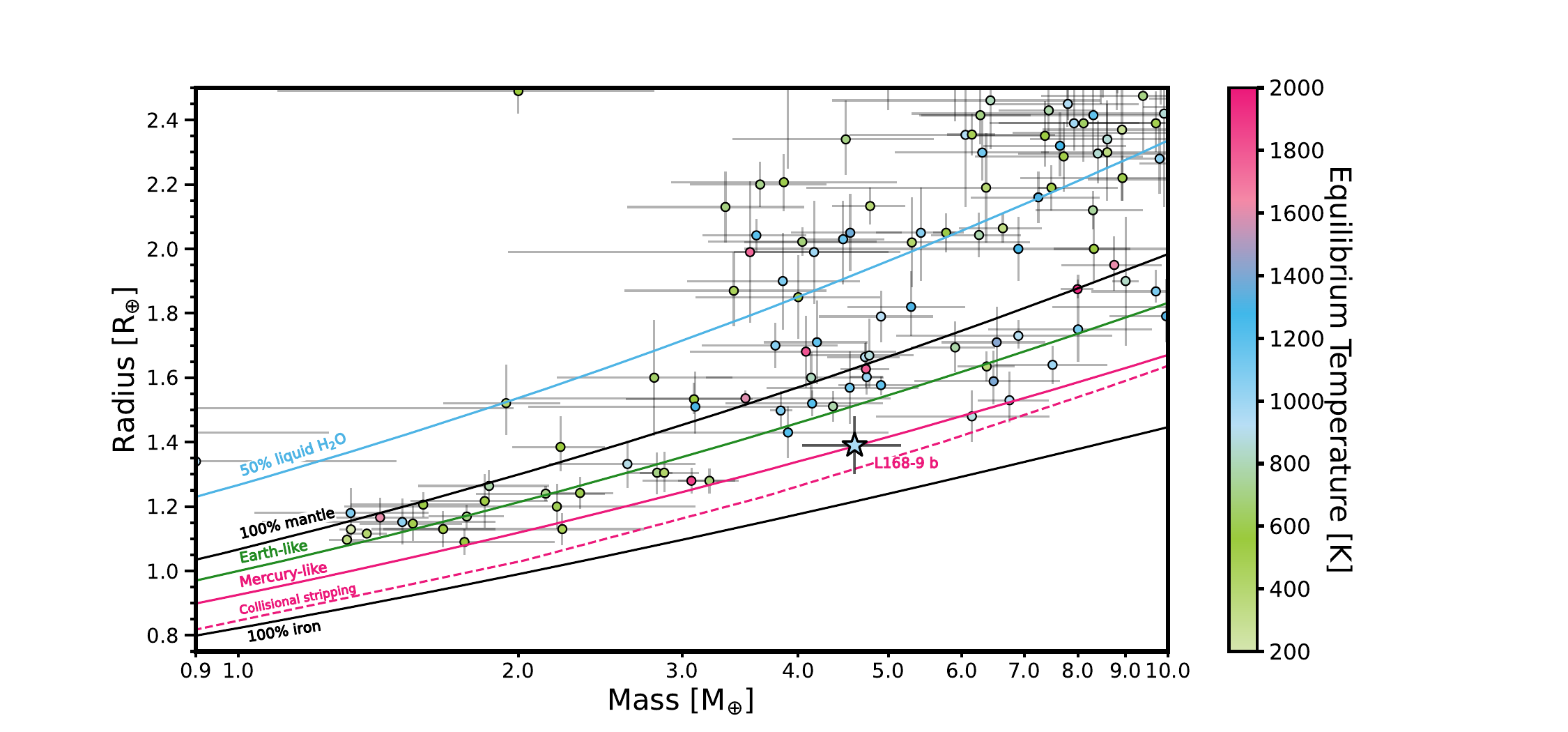}
\caption{Mass-radius diagram for small planets (points) color coded by planetary equilibrium temperature compared to theoretical curves of constant composition \citep[solid lines][]{Zeng2016}. The pink dashed line represents the smallest radius a rocky planet can have due to loss of its mantle by collisional stripping by planetesimals, based on the model of \cite{Marcus2010}. The measured mass and radius of L168-9\,b are consistent with a Mercury-like bulk composition.} 
\label{fig:mass_radius}   
\end{centering}
\end{figure*} 



Figure \ref{fig:sigma} shows the $\sigma$-significance across the opaque pressure level--metallicity parameter space given our full model grid. We compute the $\sigma$-significance from the $\chi_r^2$ values by converting the latter to p-values and then assuming normally distributed errors on the data. We find that the transmission spectrum of L\,168-9\,b rules out atmospheric compositions between 2--3$\times$ and 100$\times$ solar at $>$3$\sigma$, assuming high surface pressure ($>$1 bar), cloudless atmospheres. These metallicity bounds roughly correspond to atmospheric mean molecular weights between 2.2 and 4\,g\,mol$^{-1}$. We rule out a smaller range of metallicities for lower surface pressures and/or cloudy atmospheres, such that for opaque pressure levels $<$1 mbar all considered metallicities are allowed. 
These results are in agreement with our findings using our representative models (Figure \ref{fig:aesop_avg_tr_spec}): at low ($<$2-3$\times$ solar) metallicity, the low abundance of \ce{CO2} leads to relatively low amplitude spectral features dominated by the broad \ce{CH4} band centered at 3.3 $\mu$m; at high ($>$100$\times$ solar) metallicity, the increased atmospheric mean molecular weight reduces spectral amplitudes across all features; at intermediate metallicities, the relatively low atmospheric mean molecular weight coupled with the higher \ce{CO2} abundances leads to the appearance of the large \ce{CO2} band at 4.3 $\mu$m, which rises above the scatter of the data, requiring high altitude clouds to flatten the model to produce a good fit. 
In general we see similar results between the G395H-only and G395H+MIRI cases, except that the latter dataset is able to rule out more of the lower metallicity portion of parameter space (Figure \ref{fig:sigma}). This is likely due to the large amplitude decrease in the transit depth redward of 9 $\mu$m caused by the lack of significant molecular opacity there.

\section{Discussion} 
\label{sec:discussion}




\subsection{An Emerging Trend for Rocky Exoplanets}
The featureless transmission spectrum of L\,168-9\,b joins a growing number of similar rocky exoplanet observations with JWST (TRAPPIST-1b, \citealt{Lim2023}; LHS\,475b, \citetalias{LustigYaeger2023}; GH\,1132b, \citetalias{May23}; GJ\,486b, \citetalias{Moran2023}; TOI\,836b, \citealt{Alderson2024}; GJ\,341b, \citetalias{Kirk2024}; L98-59c, \citealt{Scarsdale_submitted}). As with this work, these studies largely conclude that their planets likely have either high mean molecular weight atmospheres, high altitude aerosols, or no atmospheres at all. Complementing these transmission studies are emission observations of rocky exoplanet daysides and/or full-orbit thermal phase curves by the Spitzer Space Telescope (LHS\,3844b, \citealt{Kreidberg2019}) and JWST (TRAPPIST-1b, \citealt{Greene23}; TRAPPIST-1c, \citealt{Zieba23}; GJ\,367b, \citealt{Zhang24_gj367b}; 55 Cnc e, \citealt{Hu24}), many of which suggest the lack of thick atmospheres capable of redistributing heat from the dayside to the nightside. All but one of these planets (55 Cnc e) orbit M dwarfs, which have high activity levels and bright, prolonged pre-main sequence phases \citep{Luger2015,Schaefer16,KrissansenTotton22}. The lack of spectral features in either emission or transmission has therefore been interpreted as the result of catastrophic stripping of light atmospheric gases (e.g., H$_2$/He) from these planets by their active host stars. After this process, high mean molecular weight atmospheres remain or no significant atmospheres at all.

\subsection{Photevaporation Modeling}\label{sec:loss}

\begin{figure*}
\begin{centering}
\includegraphics[width=0.8\textwidth]{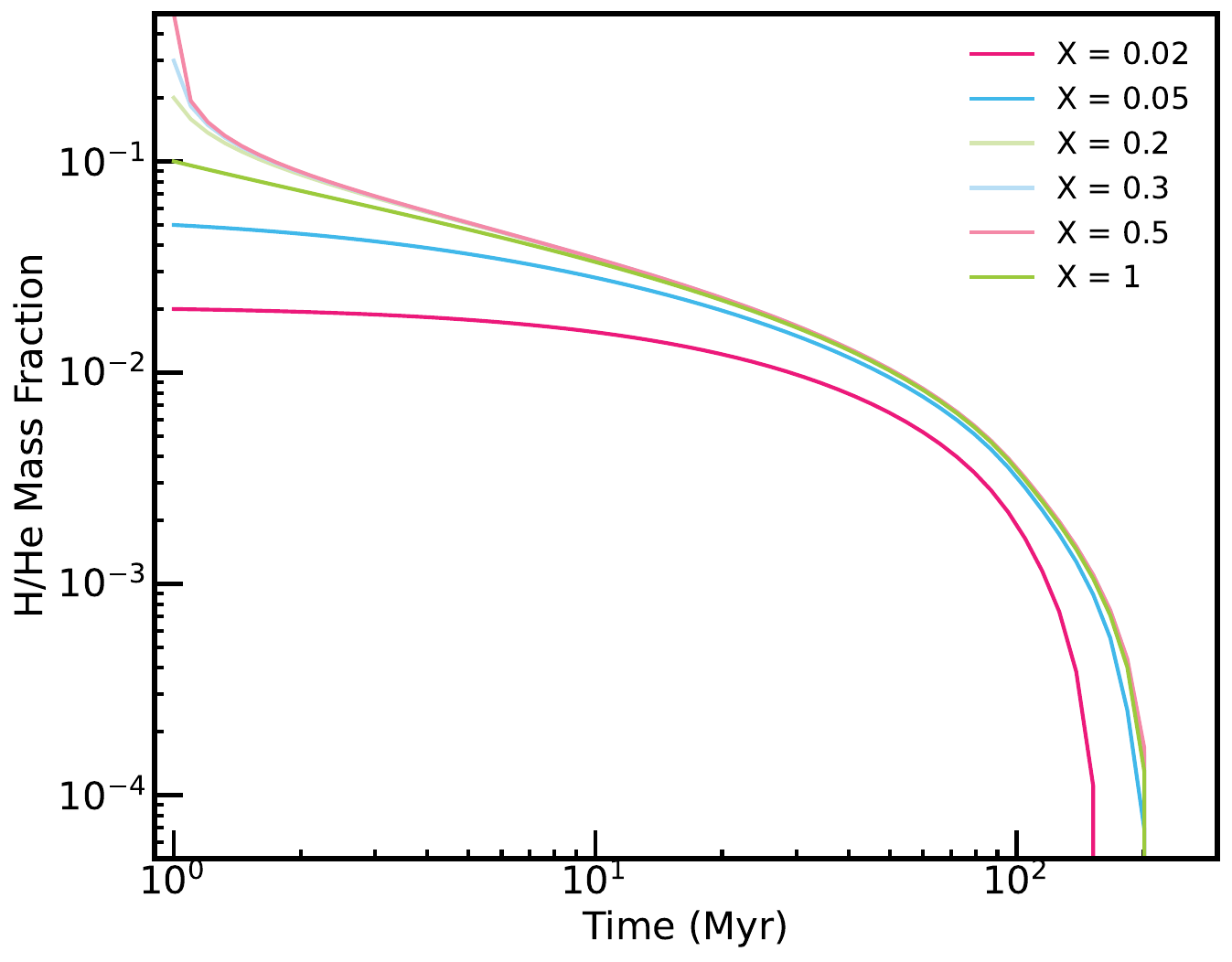}
\caption{Photoevaporation model based on \citealt{Rogers21}, adapted to the L\,168-9 system for initial H/He atmospheric mass fractions of 2\% (dark pink), 5\% (light pink), 20\% (dark blue),  30\% (light blue), 50\% (dark green), and 100\% (light green). For a large range of potential initial atmospheric mass fractions, L\,168-9 cannot retain an atmosphere beyond the first 200 Myr of its 3 Gyr lifetime.} 
\label{fig:RO_models}   
\end{centering}
\end{figure*} 

Our results for L\,168-9\,b stand out from previous works in that our data allows for near-solar metallicity atmospheres (Figure \ref{fig:sigma}). However, L\,168-9\,b's high bulk density -- greater than that of Earth (see Figure \ref{fig:mass_radius}) -- likely rules out any significant low mean molecular weight envelope. In addition, the high irradiation level of the planet is likely to result in very short lifetimes for light gases. To quantify this, we ran atmospheric escape models based on the \citet{Rogers21} photoevaporation model. We simulated planets with a core mass of 4.6$\,$M$_\oplus$ and initial atmospheric mass fractions of 0.02, 0.05, 0.1, 0.2, 0.3, 0.5, and 1.0$\times$ the core mass. We considered the evolution of the stellar XUV flux from \citet{Owen2017}, which consists of a constant, saturated XUV flux, $L_{\rm sat}$, up to the saturation time $t_{\rm sat}$, followed by a power-law decrease in time:

\begin{equation}
 L_{\rm XUV} = \begin{cases} 
          L_{\rm sat} & t < t_{\rm sat} \\
          L_{\rm sat} \left ( \frac{t}{t_{\rm sat}} \right )^{-1-a_0} & t \geq t_{\rm sat} \\
       \end{cases}
\end{equation}

\noindent where $L_{\rm sat}$ is given by

\begin{equation}
 L_{\rm sat} = 10^{-3.5}L_{\odot} \left ( \frac{M_{\star}}{M_{\odot}} \right)
\end{equation}

\noindent Following \citet{Rogers21}, we used a $t_{\rm sat}$ value of 100 Myr and set $a_0$ to 0.5, which are the appropriate values for Sun-like stars. $L_{XUV}$ for M dwarfs is similar to that of Sun-like stars \citep{Shkolnik2014}, so we can appropriately adopt these values here.
However, $t_{\rm sat}$ is longer for M dwarfs, and could reach $\sim$1 Gyr \citep{Wright11,Pineda2021}. As such, our computed envelope lifetimes are more of an upper limit. Indeed, we find that any significant primordial H$_2$/He atmosphere on L\,168-9\,b should be removed within 200 Myr (as shown in Figure \ref{fig:RO_models}), which is far less than the estimated age of the system \citep[$\sim$3 Gyr;][]{Engle2023}. 

These results suggest that, although H$_2$/He-dominated atmospheres are consistent with our data (given the scatter in the transmission spectrum), it is highly unlikely that such an atmosphere actually exists on L\,168-9\,b. We are thus left with the possibility of either a high mean molecular weight atmosphere with or without high altitude aerosols or no atmosphere at all -- consistent with previous studies of rocky exoplanets, particularly those orbiting M dwarfs. 





\subsection{Prospects for Future Atmospheric  Characterization}

\begin{figure*}[ht!]
\begin{centering}
\includegraphics[width=0.92\textwidth]{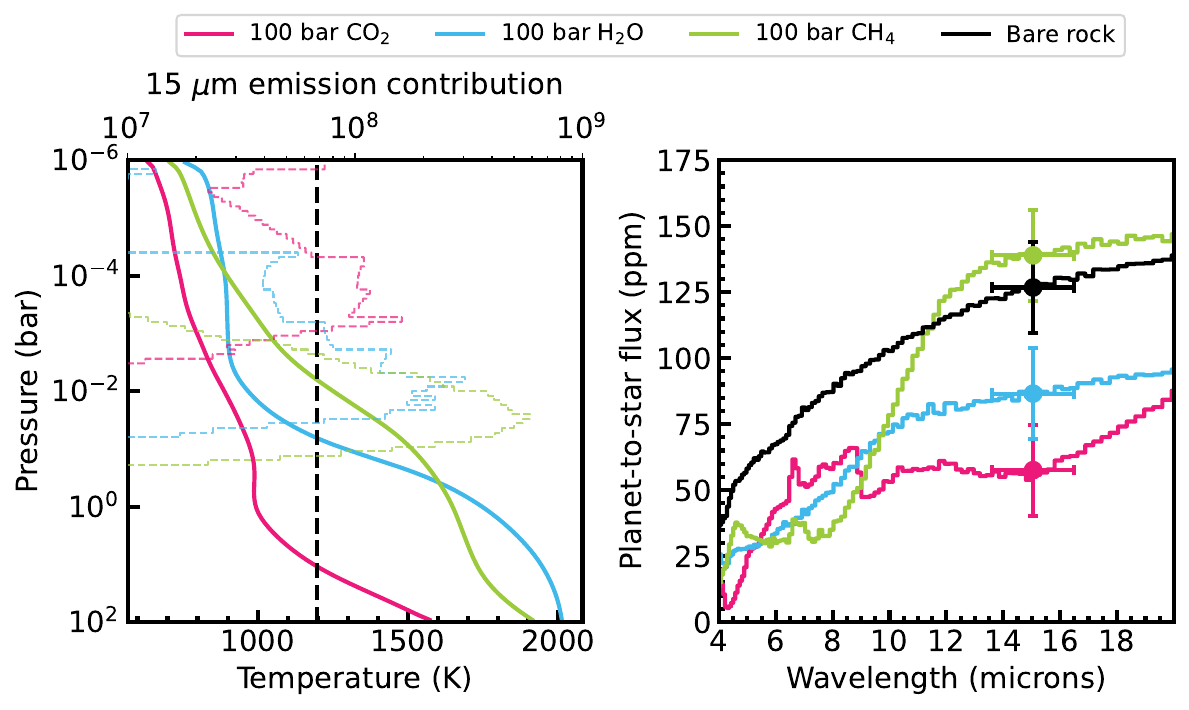}
\caption{\textit{Left}: Solid lines are the predicted temperature-pressure profiles from {\tt clima} for the 100 bar CO$_{2}$ (pink), 100 bar H$_{2}$O (blue), and 100 bar CH$_{4}$ (green) atmospheric scenarios for L168-9\,b, compared to the expected dayside emission temperature of a bare rock (1200\,K assuming an albedo of 0.1; black vertical dashed line). The colored dashed curves are the 15 $\mu$m emission contribution function (Equation 4 from \citealt{Lothringer2018}) for each atmospheric scenario. \textit{Right}: Simulated JWST/MIRI eclipse depths at 15\,$\mu$m for three eclipse observations (points) based on each scenario. The bare rock case is represented by a 1200 K blackbody. } 
\label{fig:miri_eclipses}   
\end{centering}
\end{figure*} 

Distinguishing between the high mean molecular weight atmosphere and no atmosphere scenarios for L\,168-9\,b is the next logical step for characterizing this enigmatic world, which could shed light on its initial volatile budget, external volatile delivery rate, and outgassing history \citep{Kane20,Lichtenberg22,KrissansenTotton23}. To this end, secondary eclipse observations at 15\,$\mu$m with JWST/MIRI are emerging as a powerful tool to explore the presence of an atmosphere on M dwarf rocky planets \citep{Redfield24}. We therefore explored the feasibility of 15\,$\mu$m MIRI eclipse observations for L\,168-9\,b for end-member atmospheric scenarios consistent with our single-gas forward models (\S \ref{sec:1D_models}). 

We used {\tt clima}, the 1D climate model from \citet{Wogan2023}\footnote{\url{https://github.com/Nicholaswogan/clima}}, and the day-night heat redistribution parameterization described in \citet{Koll2022} to estimate the dayside emission from 100 bar atmospheres composed of CO$_2$, H$_2$O and CH$_4$. The left panel of Figure \ref{fig:miri_eclipses} shows predicted P-T profiles, compared to the expected dayside emission temperature of a bare rock (1200\,K), assuming a 0.1 bond albedo motivated by that of Mercury.\footnote{An albedo of $\sim 0.19$ is a plausible upper bound for a bare rock, assuming an ultramafic surface \citep{Mansfield2019}.} The right panel of Figure \ref{fig:miri_eclipses} gives the resultant emission spectrum for each scenario with synthetic 15\,$\mu$m MIRI photometry observations computed with the JWST Exposure Time Calculator \citep{Pontoppidan2016}. Each synthetic data point assumes three visits, has 17 ppm precision, and ignores random Gaussian noise (i.e., the synthetic data are centered on truth). 

We find that the 100 bar CO$_2$ scenario has a relatively cold upper atmosphere because CO$_2$ emits efficiently in the infrared and is fairly transparent to the incident M-dwarf radiation. The result is a small $\sim$60\,ppm dayside emission signal at 15\,$\mu$m, which could be distinguished from a bare rock (127\,ppm emission) at $\sim$4$\sigma$ confidence for our assumed data quality. On the other hand, the H$_2$O and CH$_4$ cases absorb near-infrared starlight leading to warm upper atmospheres with large dayside emission fluxes that are more comparable to that of a bare rock \citep{Lincowski2024}. 15\,$\mu$m emission from an H$_2$O-dominated atmosphere is 2.3$\sigma$ smaller than the bare rock scenario, while a CH$_4$ atmosphere has a dayside flux within 1$\sigma$ of the bare rock case. However, with the rapid loss of hydrogen to space discussed in Section \ref{sec:loss}, it is perhaps unlikely for atmospheres rich in H-bearing species (e.g., steam or \ce{CH4}) to persist over the 3 Gyr lifetime of this planet.

\section{Summary \& Conclusions} 
\label{sec:summary}

We assembled the complete near- to mid-infrared transmission spectrum of the rocky planet L\,168-9\,b using three transit observations with NIRSpec/G395H, combined with mid-infrared MIRI/LRS commissioning data \citep{Bouwman23}. Based on the 3--12 $\mu$m transmission spectrum of this target observed with JWST, our key results on this planet's atmospheric properties are summarized below.  

\begin{itemize}
    \item The near- to mid-infrared transmission spectrum of L\,168-9\,b shows a lack of discernible atmospheric features (Figure \ref{fig:aesop_avg_tr_spec}). Fits to 1D thermochemical equilibrium forward models reveal that the data prefer single gas models and a flat line model over 25--100$\times$ solar models, while near-solar models possess intermediate goodness of fit, likely due to the lack of high amplitude spectral features at the wavelengths of interest. These results suggest a high atmospheric mean molecular weight ($>$4 g mol$^{-1}$), the presence of high altitude aerosols, or the lack of an atmosphere altogether (Figure \ref{fig:sigma}). 
    \item While L\,168-9\,b's high bulk density (Figure \ref{fig:mass_radius}) already makes low mean molecular weight atmospheres unlikely, we also ran atmospheric escape models for a range of initial atmospheric mass fractions (2--100\%) to evaluate whether small amounts of light gasses can remain on the planet. Based on these models, however, any significant primordial H$_{2}$/He atmosphere on L\,168-9\,b would be removed within 200 Myr (Figure \ref{fig:RO_models}). 
    \item 15\,$\mu$m MIRI eclipse observations could provide a discerning lens to distinguish between the high mean molecular weight and no atmosphere scenarios for L\,168-9\,b. 
    With three eclipse observations, we should be able to confidently identify a CO$_{2}$-dominated atmosphere if one exists, while emission from atmospheres of other compositions would be more challenging to distinguish from the bare rock scenario.
    
\end{itemize}

\section*{Acknowledgments}
This work is based on observations made with the NASA/ESA/CSA James Webb Space Telescope. The data were obtained from the Mikulski Archive for Space Telescopes at the Space Telescope Science Institute, which is operated by the Association of Universities for Research in Astronomy, Inc., under NASA contract NAS 5-03127 for JWST. The specific observations analyzed can be accessed via \dataset[10.17909/q5hp-7b93]{https://doi.org/DOI}. These observations are associated with program \#2512. Support for program \#2512 was provided by NASA through a grant from the Space Telescope Science Institute, which is operated by the Association of Universities for Research in Astronomy, Inc., under NASA contract NAS 5-03127. This work benefitted from the 2022 and 2023 Exoplanet Summer Program in the Other Worlds Laboratory (OWL) at the University of California, Santa Cruz, a program funded by the Heising-Simons Foundation. This work is funded in part by the Alfred P. Sloan Foundation under grant G202114194.

Co-author contributions are as follows: 
M.K.A. led the data analysis and write-up of this study. P.G. led the modeling efforts. J.I.A.R. and N.L.W. provided additional independent reductions. N.F.W. A.A., and A.D. aided in the theoretical interpretation of the data, respectively contributing climate models and simulated MIRI data, interior modeling, and photoevaporation models. All authors read and provided comments \& conversations that greatly improved the quality of the manuscript.


\bibliography{main}
\bibliographystyle{aasjournal}

\end{document}